\documentclass[prxquantum,preprint,superscriptaddress,longbibliography,amsmath,amssymb]{revtex4-2}
\usepackage{graphicx}
\graphicspath{{./figures/}}

\usepackage{booktabs}
\usepackage{siunitx}
\usepackage{braket}
\usepackage[colorlinks=true, citecolor=blue, linkcolor=blue, urlcolor=blue]{hyperref}
\usepackage[capitalize]{cleveref}
\usepackage[margin=2cm]{geometry}

\begin{document}

\title{Extraction erases the gain: cavity loss bounds dynamical-Casimir metrology at ultrastrong coupling}

\author{Juste Deuyekbe}
\email{justedeuyekbe@gmail.com}
\affiliation{Department of Physics, Faculty of Science, University of Ngaoundere, P.O. Box 454, Ngaoundere, Cameroon}

\author{P. Djorw\'e}
\email{djorwepp@gmail.com}
\affiliation{Department of Physics, Faculty of Science, University of Ngaoundere, P.O. Box 454, Ngaoundere, Cameroon}
\affiliation{Stellenbosch Institute for Advanced Study (STIAS), Wallenberg Research Centre at Stellenbosch University, Stellenbosch 7600, South Africa}

\author{Nsangou Mama}
\email{mnsangou@yahoo.com}
\affiliation{Department of Physics, Faculty of Science, University of Maroua, P.O. Box 46, Maroua, Cameroon}

\author{S. G. Nana Engo}
\email{serge.nana-engo@facsciences-uy1.cm}
\thanks{Corresponding author.}
\affiliation{Department of Physics, Faculty of Science, University of Yaounde I, P.O. Box 812, Yaounde, Cameroon}

\begin{abstract}
Ultrastrong light--matter coupling dresses the vacuum with virtual photons whose response to the coupling strength exceeds that of any coherent probe, yet these photons never leave the cavity and cannot be detected. A parametric drive at the dynamical-Casimir resonance converts them into real photons that a detector can collect. We show that this conversion is self-limiting: converting the virtual photons of an ultrastrongly coupled probe into real, extractable photons opens the cavity port that removes them, so the retained metrological gain is set by the accumulated cavity loss rather than by the pure-state ceiling. In the driven quantum Rabi model the pure-state quantum Fisher information on the coupling reaches $1.7\times10^{5}$ times the coherent baseline; at a cavity linewidth of $246$~MHz, typical of coplanar-waveguide flux-qubit circuits, master-equation dynamics retain $0.96\%$ of it. In a 32-qubit Dicke model the retained fraction is $2.6\%$ at the largest photon basis computed. Eighteen lossy configurations from both models fall on one curve of the accumulated loss, the product of linewidth and interrogation time, which halves the retained information at $0.13$. The extraction cost, not the lossless ceiling, is the figure of merit for any protocol that reads bound vacuum correlations through a lossy port.
\end{abstract}

\maketitle

\section{Introduction}
\label{sec:intro}

A quantum probe is only as useful as the information that leaves it. The quantum Fisher information (QFI) bounds the precision with which a parameter can be estimated from a state~\cite{BraunsteinCaves1994,Paris2009,Giovannetti2004,Giovannetti2011,Pezze2018}, but the bound is reached only by a measurement on the degrees of freedom that carry the parameter. In every platform---photonic, phononic, magnonic, or atomic---those degrees of freedom must cross a lossy interface to reach a detector~\cite{GardinerCollett1985,Escher2011,Kolodynski2010,Demkowicz2012,Zhou2021}. Ultrastrong light--matter coupling (USC) makes this tension acute~\cite{Ciuti2005,Ashhab2010,Niemczyk2010,Yoshihara2017,FornDiaz2017,Kockum2019,FornDiaz2019}. Its counter-rotating terms dress the ground state with virtual photons whose sensitivity to the coupling strength exceeds that of a coherent probe~\cite{DeLiberato2017,Hotter2025}, but virtual photons produce no outgoing flux~\cite{DeLiberato2017,Giannelli2024}: the gain is real and unreadable. A parametric modulation at the dynamical-Casimir (DCE) resonance converts the dressing into real radiation~\cite{Nation2012,Stassi2013,Garziano2013,Macri2018,Dodonov2020}, as demonstrated in superconducting circuits~\cite{Johansson2009,Wilson2011,Lahteenmaki2013}. Here we show that this conversion is self-limiting. Making the virtual photons of an ultrastrongly coupled probe real exposes them to the cavity port that removes them, so the retained metrological gain follows the accumulated cavity loss rather than the pure-state ceiling.

The question sits between three bodies of work that have not been joined. Virtual-state and critical metrology show that a dressed or near-critical ground state is metrologically rich but leave the information inside the system~\cite{Hotter2025,Hotter2024,Garbe2020,Chu2021}. Studies of DCE metrology benchmark the generated field against classical states of equal photon number in a closed or short-dwell encoding~\cite{Schelb2026}. Lossy-metrology theory bounds the information that survives a fixed channel applied after encoding~\cite{Escher2011,Kolodynski2010,Demkowicz2012,Zhou2021,Kurdzialek2025}, and Floquet engineering protects it against emitter-local baths~\cite{Bai2023}. None of these follows a driven USC probe whose loss acts while the parameter is being encoded, through the same port that later delivers the signal, at a fixed device linewidth. Dissipation at ultrastrong coupling also requires care, because bare and dressed descriptions of the bath differ~\cite{Beaudoin2011,Settineri2018}.

We close this gap with exact-diagonalization and master-equation calculations of the quantum Rabi and $N=32$ Dicke models under a modulated coupling. The static dressed vacuum carries a sixfold gain that never reaches the port. The DCE drive raises the pure-state QFI by five orders of magnitude, and the port removes more than \qty{99}{\percent} of it at a coplanar-waveguide linewidth. Eighteen lossy configurations from both models collapse onto an empirical extraction law in the accumulated loss $x=\kappa\tau_{\rm int}$, with the retained information halved at $x_c\approx\num{0.13}$, and the collapse falls below the bare transmissivity of the port because loss reshapes the parameter-dependent trajectory during the drive. We translate these results into a four-step decision protocol for driven USC sensors.

\section{Model and figures of merit}
\label{sec:methods}

\begin{figure}[htbp]
\centering
\includegraphics[width=0.6\columnwidth]{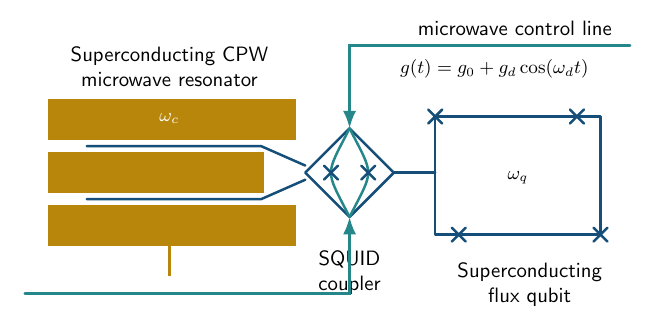}
\caption{Probe architecture (schematic, no data). A CPW resonator (frequency $\omega_c$, linewidth $\kappa$) couples to a flux qubit (splitting $\omega_q$) through a SQUID-mediated tunable coupler. Flux modulation of the coupler yields $g(t)=g_0+g_d\cos(\omega_d t)$; the output port carries the emitted field to the detector.}
\label{fig:setup}
\end{figure}

A single flux-tunable element implements both probes studied here: a superconducting coplanar-waveguide (CPW) resonator couples to a flux qubit through a SQUID-mediated coupler whose coupling $g(t)=g_0+g_d\cos(\omega_d t)$ is static for $g_d=0$ and parametrically driven otherwise (\cref{fig:setup})~\cite{Niemczyk2010,Yoshihara2017,Kockum2019}.

\paragraph{Hamiltonians.} The single-qubit quantum Rabi Hamiltonian is ($\hbar=1$)
\begin{equation}\label{eq:main_rabi}
\hat{H}_{\text{Rabi}} = \frac{\omega_q}{2}\hat{\sigma}_z + \omega_c \hat{a}^\dagger \hat{a} + g(t)\, \hat{\sigma}_x (\hat{a} + \hat{a}^\dagger),
\end{equation}
with normalized coupling $\eta = g_0/\omega_c$, and the $N$-qubit Dicke Hamiltonian is
\begin{equation}\label{eq:main_dicke}
\hat{H}_{\text{Dicke}} = \omega_c \hat{a}^\dagger \hat{a} + \omega_q \hat{J}_z + \frac{g(t)}{\sqrt{N}}\, \hat{J}_x (\hat{a} + \hat{a}^\dagger),
\end{equation}
where $\hat{J}_{x,z} = \tfrac12\sum_{i=1}^N \hat{\sigma}_{x,z}^{(i)}$. Its thermodynamic superradiant threshold is $\eta_c = \tfrac12\sqrt{\omega_q/\omega_c} = \num{0.5}$ at resonance~\cite{Gilmore1976Pers,Emary2003,Kirton2019}. A flux bias $\Phi_{\rm ext}(t)=\Phi_0/2+\delta\Phi\cos(\omega_d t)$ on the coupler generates the modulation. We work with this effective coupling; a gauge-consistent treatment of nonadiabatic modulation~\cite{Stokes2021,DiStefano2019} is discussed in \cref{sec:limitations_B}, and a finite-dimensional similarity transformation reproduces the QFI metric to \num{5e-9} (\cref{SI-sec:si_multiparameter}). Notation is collected in \cref{SI-tab:si_notation_b,SI-tab:si_trackB_glossary}.

\paragraph{Quantum Fisher information.} For the static ground state $\ket{\Psi_G(\eta)}$ the QFI is the gap-weighted sum
\begin{equation}\label{eq:main_qfi_static}
F_Q = 4\sum_{k\neq 0}\frac{|\braket{\Psi_k|\hat{\sigma}_x(\hat{a}+\hat{a}^\dagger)|\Psi_G}|^2}{(E_k-E_0)^2},
\end{equation}
quoted relative to the coherent-state baseline $F_Q^{\rm coh}=1$~\cite{BraunsteinCaves1994,Giovannetti2011}. The variance form $4\,\mathrm{Var}[\hat\sigma_x(\hat a+\hat a^\dagger)]$ applies only to unitary encodings and overestimates the static QFI by a factor \num{3.4} at its maximum (\cref{SI-sec:si_energy_budget}). The driven state is nonstationary, and we use the fidelity form~\cite{Liu2020}
\begin{equation}\label{eq:main_qfi_dce}
F_Q(t) = 8\,\frac{1-|\braket{\Psi_\eta(t)|\Psi_{\eta+\delta}(t)}|}{\delta^2}, \qquad \delta=\num{1e-3},
\end{equation}
with the Uhlmann fidelity replacing the overlap for mixed states. Because \cref{eq:main_qfi_dce} reads a fidelity defect of order $10^{-6}$, every propagation uses absolute and relative tolerances of \num{1e-10} and \num{1e-8}; the lossless master-equation run then returns \num{0.9986} of the independently computed pure-state peak, which fixes the resolution of every retention reported below at \qty{0.1}{\percent} (\cref{SI-sec:si_loss_adjusted}).

\paragraph{Open-system dynamics.} Cavity decay enters through the Lindblad operator $\sqrt{\kappa}\,\hat a$~\cite{Gardiner2004}, propagated with QuTiP~\cite{Johansson2012} in Fock bases of $N_{\rm cav}=\numrange{6}{28}$ photons. We test the Markovian treatment against a hierarchical-equations-of-motion (HEOM) calculation~\cite{Tanimura2020,Lambert2023} in which the cavity mode couples through its position quadrature to a Drude--Lorentz bath (\cref{SI-sec:si_cavity_heom}); at $\kappa/\omega_c=\num{1e-4}$ the two treatments coincide to a state infidelity below \num{4e-4}. Retention is the ratio $R_Q=F_Q^{\rm diss}/F_Q^{\rm ideal}$ at matched interrogation window. Crossovers of the single-qubit loss sweep are log-linear interpolations between sampled rows. The empirical extraction law is a least-squares fit with bootstrap \qty{95}{\percent} intervals (\cref{SI-sec:si_extraction}). Multilevel leakage is screened with three-level transmon-like, flux-qubit-like and fluxonium-like spectra~\cite{Koch2007,Yan2016,Manucharyan2009} against a ceiling of \qty{1}{\percent} population outside the qubit subspace over \num{40} drive periods (\cref{SI-sec:si_multilevel}). Convergence diagnostics are in \cref{SI-sec:si_convergence}.

\paragraph{Extraction inequality and port budget.} For a completely positive trace-preserving map $\Lambda$ applied after encoding, monotonicity of the QFI gives $F_Q(\Lambda(\rho_\theta))\le F_Q(\rho_\theta)$~\cite{BraunsteinCaves1994,Giovannetti2011}, so $R_Q\le1$. For a Gaussian displacement probe sent through pure attenuation of transmissivity $\eta_{\rm t}$ the information composes affinely, $F_Q^{\rm out}=\eta_{\rm t} F_Q^{\rm in}+(1-\eta_{\rm t})F_{\rm vac}$~\cite{Escher2011,Kolodynski2010,Demkowicz2012}. The transmissivity that a single-mode port assigns to the field leaving the cavity over an interrogation time $\tau_{\rm int}$ is
\begin{equation}
\label{eq:eta_eff}
\eta_{\rm eff}=\frac{\kappa_{\rm ext}}{\kappa}\,\frac{1-e^{-\kappa\tau_{\rm int}}}{\kappa\tau_{\rm int}}\,e^{-\gamma_q\tau_{\rm int}},\qquad \gamma_q=\frac{1}{T_1}+\frac{1}{T_\phi}.
\end{equation}
In the driven protocol loss acts during the encoding, so the lossy and lossless runs are two different parameter-dependent evolutions and the inequality is a reference rather than a theorem; every computed row nevertheless satisfies $R_Q<1$ and lies below the bare envelope $e^{-\kappa\tau_{\rm int}}$, and \cref{sec:extraction} explains why.

\section{Results}
\label{sec:results}

Making the virtual photons of an ultrastrongly coupled probe real exposes them to the cavity port that removes them, so the retained metrological gain follows the accumulated cavity loss rather than the pure-state ceiling. We establish this in four steps. The static dressed vacuum holds a gain that never reaches the port (\cref{sec:scaling}). The DCE drive makes that gain extractable and radiating at once (\cref{sec:dce}). The radiation cost then collapses onto one law of the accumulated loss, behind ceilings already set by readout and nuisance parameters (\cref{sec:extraction,sec:readout}). The collective Dicke probe meets the same barrier (\cref{sec:dicke,sec:dicke_dce}). \Cref{sec:optimization} turns these results into operating rules.

\subsection{Static dressing is a non-emissive resource}
\label{sec:scaling}

\begin{figure}[htbp]
    \centering
    \includegraphics[width=.6\columnwidth]{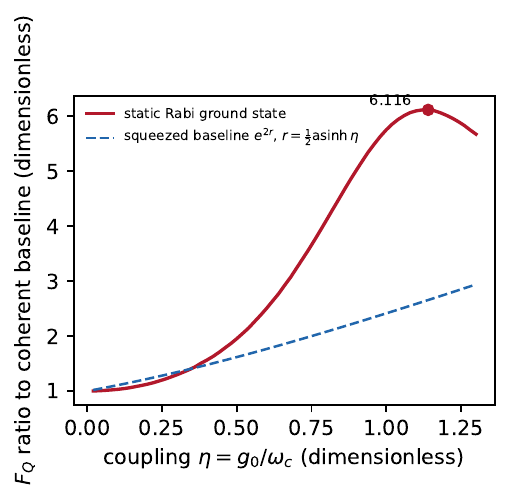}
    \caption{The dressed vacuum is a better probe of the coupling than a coherent or squeezed state. Static Rabi ground-state QFI (relative to the coherent baseline) versus $\eta$, reaching $\num{6.12}$ at the sampled maximum $\eta=\num{1.14}$, compared with the squeezed-state reference $e^{2r}$, $r=\tfrac12\sinh^{-1}\eta$, which it exceeds for $\eta>\num{0.3}$.}
    \label{fig:qfi_comparison}
\end{figure}

The dressed Rabi ground state is a better probe of the coupling than a coherent state, and its advantage grows with $\eta$ up to a sixfold maximum. \Cref{fig:qfi_comparison} shows the QFI rising from the perturbative law $F_Q-1\approx3\eta^2$ (coefficient \num{3.0} as $\eta\to0$, \num{3.8} at $\eta=\num{0.5}$; \cref{SI-sec:si_energy_budget}) to $F_Q=\num{6.12}$ at the sampled maximum $\eta=\num{1.14}$, where the ground state hosts $n_{\rm virt}=\num{1.03}$ virtual photons (\cref{tab:trackB_usc_scaling}). Beyond this point the QFI declines as the two parity-split minima of the deep-strong-coupling (DSC) ground state approach degeneracy~\cite{Ashhab2010,Kockum2019,Girvin2019Schr}. For $\eta>\num{0.3}$ the dressed vacuum outperforms the squeezed-state reference $e^{2r}$ with $r=\tfrac12\sinh^{-1}\eta$.

\begin{table}[hbt!]
\centering
\caption{Static Rabi ground-state QFI relative to the coherent baseline, with the virtual photon number $n_{\rm virt}=\braket{\Psi_G|\hat a^\dagger\hat a|\Psi_G}$. The maximum of the 65-point sweep lies at the sampled point $\eta=\num{1.14}$.}
\label{tab:trackB_usc_scaling}
\begin{tabular}{c c c}
\toprule
$\eta$ & $n_{\text{virt}}$ & $F_Q/F_Q^{\rm coh}$ \\
\midrule
\num{0.10} & \num{0.0025} & \num{1.03} \\
\num{0.80} & \num{0.3282} & \num{4.11} \\
\num{1.00} & \num{0.6814} & \num{5.75} \\
\num{1.14} & \num{1.0339} & \num{6.12} \\
\num{1.30} & \num{1.5084} & \num{5.67} \\
\bottomrule
\end{tabular}
\end{table}

\begin{figure}[htbp]
    \centering
    \includegraphics[width=.9\columnwidth]{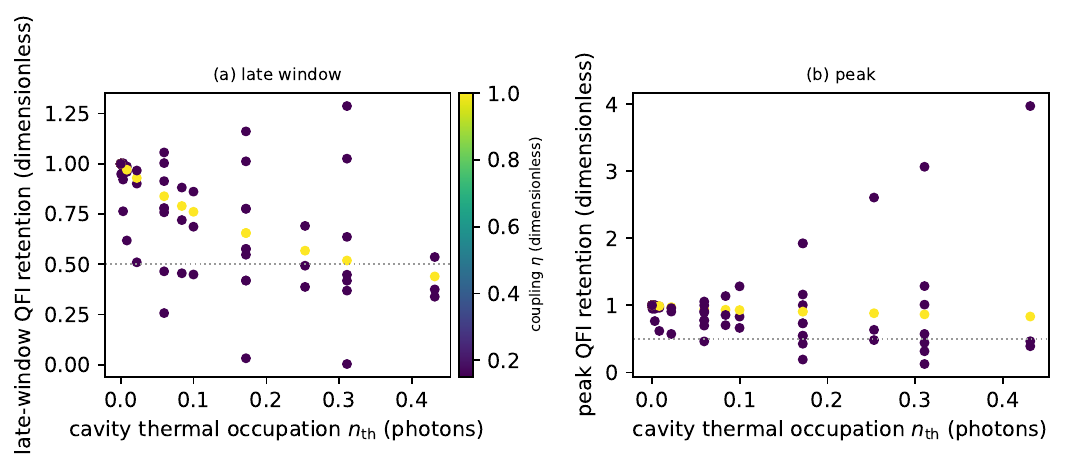}
    \caption{Thermal occupation, not the port, limits the static gain. (a) Late-window QFI retention of \num{98} finite-temperature configurations against the cavity thermal occupation $n_{\rm th}$, colored by $\eta$. (b) The same configurations scored on their peak QFI. Dotted line: \qty{50}{\percent} retention, cleared by \num{82} configurations in (a) and \num{89} in (b).}
    \label{fig:no_go_boundaries}
\end{figure}

This gain sits in the curvature of the dressed ground state with respect to $\eta$, not in any outgoing field. The virtual photons that produce it are bound to the qubit and do not populate the output mode~\cite{DeLiberato2017,Giannelli2024}, so the static QFI is a non-emissive resource: it sets the ceiling for any readout channel but supplies no channel of its own. The same binding shelters it from the port. Its fragility lies instead in thermal mixing of the dressed vacuum (\cref{fig:no_go_boundaries}): in a finite-temperature screen of \num{98} configurations, \num{82} retain at least half of their late-window QFI and \num{89} at least half of their peak QFI as the cavity thermal occupation $n_{\rm th}=(e^{\hbar\omega_c/k_BT}-1)^{-1}$ grows~\cite{Clerk2010}.

Reading the static gain therefore requires a mechanism that turns bound dressing into radiation. That mechanism is the DCE drive.

\subsection{The DCE drive makes the gain extractable and radiating}
\label{sec:dce}

\begin{figure}[htbp]
    \centering
    \includegraphics[width=.9\columnwidth]{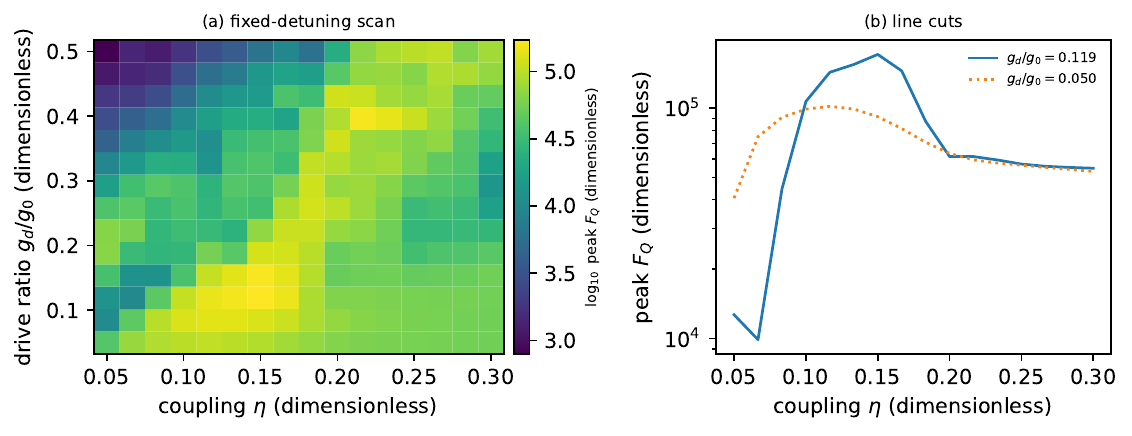}
    \caption{A narrow resonance in drive amplitude sets the driven gain. (a) Pure-state peak QFI in the $(\eta, g_d/g_0)$ plane at $\omega_q/\omega_c=\num{1.027}$ and $\omega_d/\omega_c=\num{2.077}$; the maximum $\num{170903}$ lies at $\eta=\num{0.150}$, $g_d/g_0=\num{0.11923}$. (b) Line cuts at fixed $g_d/g_0$.}
    \label{fig:dce_optimum}
\end{figure}

A parametric drive near the dressed two-photon resonance turns the static sixfold gain into a five-orders-of-magnitude gain, because it pumps the vacuum into the excited manifolds whose response to $\eta$ is far larger than the ground-state curvature~\cite{Akbari2025,Larson2024}. \Cref{fig:dce_optimum} maps the pure-state peak QFI over the $(\eta,g_d/g_0)$ plane at fixed $\omega_q/\omega_c=\num{1.027}$ and $\omega_d/\omega_c=\num{2.077}$. The maximum, $\num{1.7e5}$ times the coherent baseline, lies at $\eta=\num{0.150}$ and $g_d/g_0=\num{0.11923}$, where the static value is $\num{1.07}$ (\cref{tab:trackB_dce}).

\begin{table}[hbt!]
\centering
\caption{Pure-state DCE QFI at the maximum of the $16\times14$ fixed-detuning scan over $(\eta, g_d/g_0)$, single qubit, relative to the coherent baseline.}
\label{tab:trackB_dce}
\begin{tabular}{l c}
\toprule
Quantity & Value \\
\midrule
Peak QFI (exact scan maximum) & $\num{170903}$ \\
Last-window mean & $\num{1.3e5}$ \\
Full-window mean & $\num{4.7e4}$ \\
Peak photon number $n_{\text{peak}}$ & \num{9.3} \\
\midrule
\multicolumn{2}{l}{\footnotesize Operating point: $\eta=\num{0.150}$, $g_d/g_0=\num{0.11923}$, $\omega_q/\omega_c=\num{1.027}$, $\omega_d/\omega_c=\num{2.077}$.} \\
\bottomrule
\end{tabular}
\end{table}

\begin{figure}[htbp]
    \centering
    \includegraphics[width=.8\columnwidth]{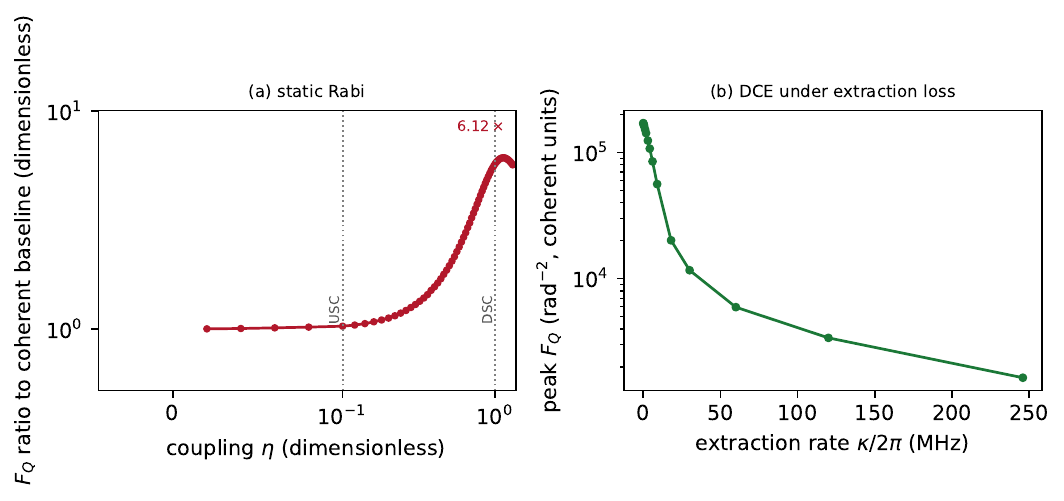}
    \caption{The driven gain exceeds the static one by four orders of magnitude and then drains through the port. (a) Static Rabi QFI relative to the coherent baseline versus $\eta$ across the USC and DSC regimes, with its sampled maximum \num{6.12}. (b) Peak DCE QFI at the single-qubit operating point versus the extraction rate $\kappa/2\pi$, from the lossless value \num{1.7e5} to \num{1642} at \qty{246}{\mega\hertz}.}
    \label{fig:shaking_catalysis}
\end{figure}

The resonance is narrow in amplitude, and driving harder spends pump power without buying information. The peak QFI per unit of the pump-energy proxy, $F_Q^{\rm peak}/(g_d/g_0)^2$, falls from \num{3.67e7} at $g_d/g_0=\num{0.05}$ to \num{2.45e4} at $\num{0.50}$, and the absolute peak itself turns over beyond the optimum, dropping to \num{6131} at the largest amplitude (\cref{SI-sec:si_energy_budget}). We attribute the turnover to drive-induced shifts of the dressed quasienergies away from the fixed resonance and to saturation of the two-level emitter; the weak-drive resonance is the efficient operating regime.

\begin{figure}[htbp]
    \centering
    \includegraphics[width=.98\columnwidth]{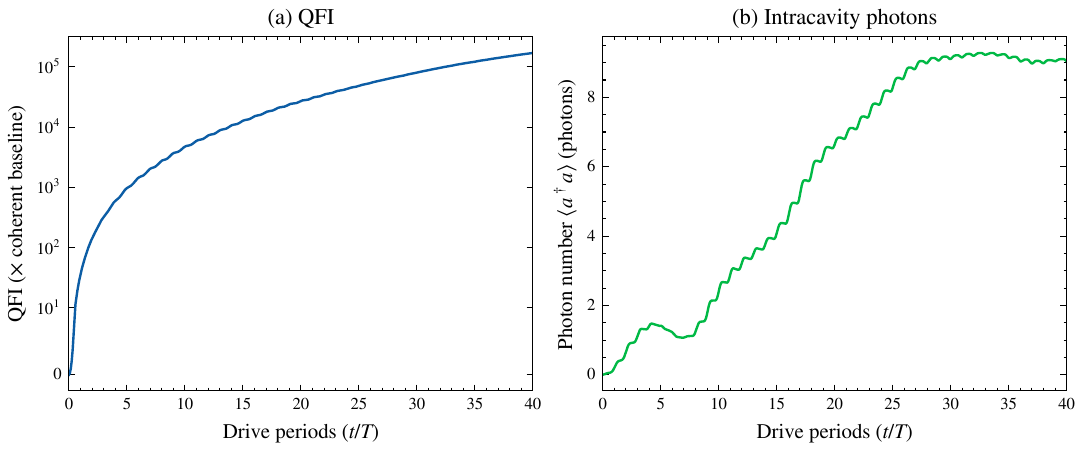}
    \caption{The driven gain is carried by real intracavity photons. (a) Lossless QFI trace at the rounded operating point $\eta=\num{0.150}$, $g_d/g_0=\num{0.119}$, $\omega_q/\omega_c=\num{1.027}$, $\omega_d/\omega_c=\num{2.077}$; the peak is \num{170492} and the last-window mean \num{123638}, close to but distinct from the scan values of \cref{tab:trackB_dce}. (b) The photon population follows the same cycle.}
    \label{fig:dce_oscillations}
\end{figure}

The gain is carried by real photons, and real photons leave. The driven state reaches $n_{\rm peak}=\num{9.3}$ intracavity photons and oscillates on a \num{34}-period cycle (\cref{fig:dce_oscillations}). Under single-photon cavity loss~\cite{Gardiner2004,GardinerCollett1985} the retained peak QFI falls monotonically across a fifteen-point sweep (\cref{SI-sec:si_loss_adjusted,SI-sec:si_kappa_sweep}): \qty{94.3}{\percent} survives at $\kappa/2\pi=\qty{0.6}{\mega\hertz}$, the retention crosses \qty{80}{\percent} at $\kappa/2\pi=\qty{2.0}{\mega\hertz}$ and \qty{50}{\percent} at \qty{5.9}{\mega\hertz}, and at the CPW operating point $\kappa/\omega_c=\num{4.1e-2}$ ($\kappa/2\pi=\qty{246}{\mega\hertz}$ for $\omega_c/2\pi=\qty{6}{\giga\hertz}$) only \qty{0.96}{\percent} remains, a QFI of \num{1642} (\cref{fig:shaking_catalysis}b). Adding a \qty{20}{\milli\kelvin} thermal drive and a conservative qubit relaxation to the physical loss changes the absolute peak by less than \qty{0.2}{\percent} (\cref{SI-sec:si_joint_row}). The port, not the bath, sets the loss.

\subsection{The extraction barrier}
\label{sec:extraction}

\begin{figure}[htbp]
    \centering
    \includegraphics[width=\columnwidth]{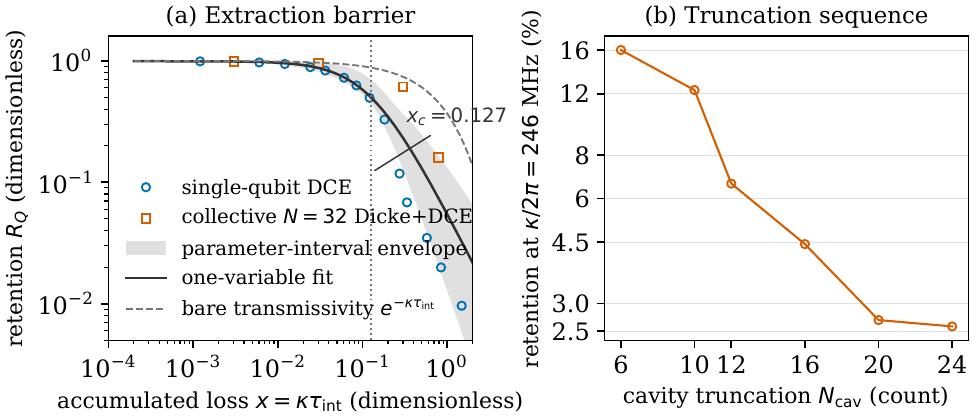}
    \caption{Both probe families collapse onto one extraction law. (a) Peak-QFI retention of the eighteen lossy configurations against the accumulated loss $x=\kappa\tau_{\rm int}$: single-qubit DCE (circles) and $N=\num{32}$ Dicke+DCE (squares). Solid line, \cref{eq:extraction_law} with $R_0=\num{0.996}$, $x_c=\num{0.127}$, $\beta=\num{1.38}$; band, envelope of the marginal \qty{95}{\percent} intervals of $x_c$ and $\beta$; dashed line, bare transmissivity $e^{-\kappa\tau_{\rm int}}$; dotted line, $x_c$. (b) Dicke+DCE retention at the CPW operating point $\kappa/\omega_c=\num{4.1e-2}$ for six cavity truncations (the axis label quotes the \qty{248}{\mega\hertz} literature linewidth; the computation uses \qty{246}{\mega\hertz}).}
    \label{fig:extraction_law}
\end{figure}

The collapse is not the attenuation of a signal the protocol already held; it is the cost of creating the signal in a mode that leaks. Two facts establish this. First, retention lies below the bare amplitude-damping envelope $e^{-\kappa\tau_{\rm int}}$~\cite{GardinerCollett1985,Escher2011} at every sampled loss, by \qty{0.7}{\percent} at the weakest and by a factor \num{24} at the CPW linewidth, because loss acting during the drive reshapes the whole parameter-dependent Floquet trajectory rather than scaling the emitted amplitude. Second, the retention of both probe families collapses onto a single function of the accumulated loss $x=\kappa\tau_{\rm int}$ (\cref{fig:extraction_law}a).

\Cref{fig:dicke_scaling} shows the static QFI of the Dicke model as a function of $N$ and $\eta$ (\cref{tab:trackB_dicke} tabulates the $\eta=\num{1.15}$ values). In the superradiant regime ($\eta \approx \num{1.15}$), the QFI exhibits collective enhancement. Fitted on the $N\le\num{18}$ data at fixed $\eta=\num{1.15}$ and $\omega_q/\omega_c=\num{1.005}$ (where the data are monotonic and not visibly saturating), the power law $F_Q \propto N^\alpha$ gives $\alpha = \num{2.216}\pm\num{0.034}$ ($R^2=\num{0.9988}$, $95\%$ CI: $[\num{2.150}, \num{2.282}]$, which excludes unity); the super-linear scaling therefore exceeds the linear expectation for independent qubits. Growth slows strongly at larger $N$: $F_Q(N=\num{27})=\num{229}\times$ and $F_Q(N=\num{33})=\num{238}\times$ (only $\sim\qty{4}{\percent}$ increase over a \qty{22}{\percent} increase in $N$), and the local growth exponent falls below unity beyond $N\approx\num{21}$. The growth nevertheless continues over the tested window rather than saturating into a tested form: an extended grid (method in \cref{SI-sec:si_finite_size}) up to $N=\num{64}$ gives $F_Q\approx\num{350}\times$, which single-plateau forms do not describe. Saturation beyond the window is not excluded. The reported exponent therefore characterizes the small-$N$ ($N\le\num{18}$) super-linear regime without extrapolating to $N\to\infty$.

\paragraph{Empirical extraction law.} The eighteen positive-loss rows of the single-qubit and $N=\num{32}$ Dicke families follow
\begin{equation}
\label{eq:extraction_law}
R_Q(x)=\frac{R_0}{1+(x/x_c)^{\beta}},\qquad x=\kappa\tau_{\rm int},
\end{equation}
with $R_0=\num{0.996}$, $x_c=\num{0.127}$ (bootstrap \qty{95}{\percent} interval \numrange{0.107}{0.205}), $\beta=\num{1.38}$ (\numrange{1.02}{2.13}) and $R^2=\num{0.919}$, with both intervals excluding zero (\cref{SI-sec:si_extraction}). Here $\tau_{\rm int}$ is fixed row by row at the maximum of the retention over the interrogation window. The saturation is unity within the fit's reach: pinning $R_0=1$, the data-processing value~\cite{BraunsteinCaves1994,Zhou2021}, returns $x_c=\num{0.126}$ and $\beta=\num{1.37}$ with unchanged $R^2$. The barrier is therefore carried entirely by the knee and the exponent. At $x=x_c\approx\num{0.13}$ half the information is gone, and beyond it retention falls as $x^{-1.4}$. Both intervals are interior to the fit domain because the single-qubit sweep brackets the knee from both sides, $x\in[\num{0.024},\num{0.18}]$. Fitted inside each family the knee separates: $x_c=\num{0.109}$ (\numrange{0.095}{0.117}) with $\beta=\num{1.73}$ for the single-qubit rows, and $x_c=\num{0.371}$ (\numrange{0.174}{4.62}) with $\beta=\num{2.17}$ for the four $N=\num{32}$ rows, at $R^2=\num{0.9928}$ and $R^2=\num{0.9959}$ against \num{0.919} pooled (\cref{SI-sec:si_extraction_families}). The two probe architectures therefore share the functional form rather than a common scaling variable: one saturating curve brackets both to within a factor \num{3} in $x_c$, and the pooled fit is a description of the union, not a universal knee.

Ranked against alternative forms on the same \num{18} rows, the saturating form beats the bare transmissivity by $\Delta\mathrm{AIC}=\num{36.5}$, a single exponential by $\Delta\mathrm{AIC}=\num{1.22}$ and the free-saturation variant by $\Delta\mathrm{AIC}=\num{1.99}$, with leave-one-out error ordering the four forms identically (\cref{SI-sec:si_extraction_selection}).

\paragraph{Loss shortens the useful interrogation.} The accumulated loss is the natural variable because loss moves the optimal readout time. In the single-qubit probe the maximum of the information rate $F_Q(t)/t$ migrates from \num{18.08} to \num{6.96} drive periods as $\kappa/\omega_c$ grows from \num{0.005} to \num{0.041}, and the intracavity occupation at that optimum falls from \num{5.08} to \num{0.76} photons (\cref{SI-sec:si_interrogation}). Across the same range the von Neumann entropy of the probe at the rate maximum stays within $S\in[\num{1.35},\num{1.60}]$~bits, an effective Hilbert-space dimension $2^{S}$ between \num{2.5} and \num{3.0}, while the peak-QFI time sits at $S\in[\num{1.80},\num{2.05}]$~bits. Coherent sensitivity builds until dissipation drives the probe to this mixedness ceiling; stronger loss reaches the ceiling sooner, and the useful window closes proportionally earlier. \paragraph{The collective probe shortens the same way.} Its optimum moves from \num{9.65} drive periods without loss to \num{6.13} at the largest loss. Because retention is always evaluated at matched window, this migration enters the numerator and not the reference.

\subsection{Readout and nuisance ceilings in front of the port}
\label{sec:readout}

The port is the last of three ceilings, and the first two are already low. A lossless probe state does not deliver its QFI to a laboratory record by itself, whether the readout is dispersive homodyne or the parametric longitudinal modulation that renders the coupling QND~\cite{Didier2015}: monitoring only the external fraction of the emitted field over a flat temporal mode of six or ten drive periods yields a homodyne classical Fisher information between $0$ and \num{0.603} of the joint-system QFI across the sampled configurations, and \num{0.0455} for the largest static case (\cref{SI-sec:si_homodyne}).

Nuisance parameters impose the second ceiling, and here the drive helps rather than hurts. In the static ground state, $\eta$ and $\omega_q/\omega_c$ deform the dressed vacuum along nearly the same direction, so their Fisher matrix is close to singular and treating $\omega_q$ as unknown erases almost all the coupling information through the Schur complement $F_{\eta\eta}-F_{\eta\omega}^2/F_{\omega\omega}$~\cite{Ragy2016,Szczykulska2016,Liu2020}. Entering the DSC regime or switching on the DCE drive lifts the degeneracy (\cref{tab:multiparameter}). Under the drive, $\eta$ sets the amplitude of the resonant pair-creation channel while $\omega_q/\omega_c$ detunes it, so the two derivatives no longer describe one deformation; in a simultaneous four-parameter estimate of $\eta$, $\omega_q/\omega_c$, $g_d/g_0$ and the drive phase the driven probe keeps \qtyrange{23}{50}{\percent} of its diagonal information (\cref{SI-sec:si_multiparameter}). The radiative barrier thus acts on a record that is already readout- and nuisance-limited, and it is the one ceiling that no choice of estimator removes.

\begin{table}[hbt!]
\centering
\caption{Nuisance-projected coupling information. Correlation $\rho_{\eta\omega}$ and condition number of the $(\eta,\omega_q/\omega_c)$ Fisher matrix, and retained fraction $F_{\eta\eta}^{\rm eff}/F_{\eta\eta}$ of the diagonal coupling information. The driven row is a four-parameter estimate (\cref{SI-sec:si_multiparameter}).}
\label{tab:multiparameter}
\begin{tabular}{l c c c}
\toprule
Probe & $\rho_{\eta\omega}$ & Condition number & Retained fraction \\
\midrule
Static, $\eta=\num{0.10}$ & \num{-0.99755} & \num{8.63e4} & \qty{0.49}{\percent} \\
Static, $\eta=\num{0.15}$ & \num{-0.99460} & \num{1.81e4} & \qty{1.08}{\percent} \\
Static DSC, $\eta=\num{1.0}$ & \num{-0.89461} & \num{175.3} & \qty{20.0}{\percent} \\
DCE-driven, $\eta=\num{0.15}$ & --- & \numrange{794}{1165} & \qtyrange{23}{50}{\percent} \\
\bottomrule
\end{tabular}
\end{table}

\subsection{Collective dressing in the Dicke model}
\label{sec:dicke}

\begin{figure}[htbp]
    \centering
    \includegraphics[width=.9\columnwidth]{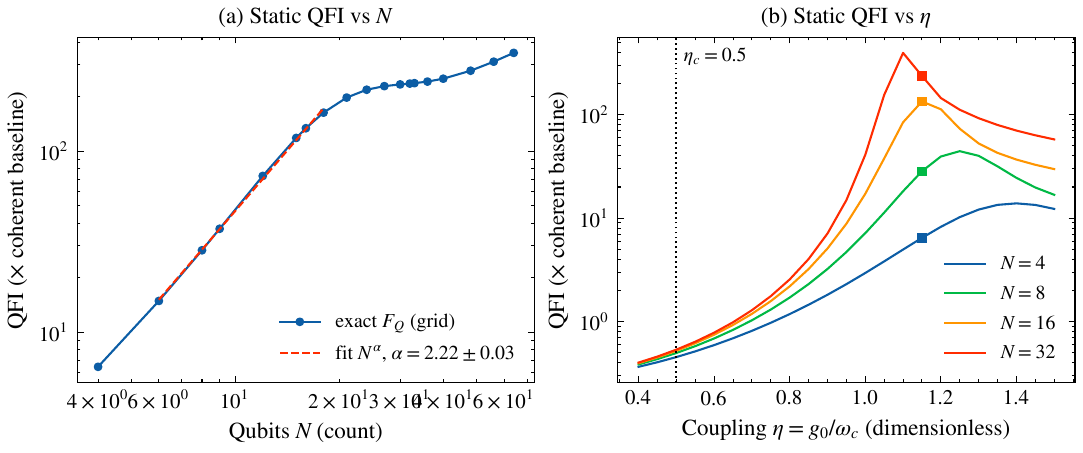}
    \caption{Collective dressing grows super-linearly at small $N$. (a) Static QFI versus $N$ at $\eta=\num{1.15}$, $\omega_q/\omega_c=\num{1.005}$; the fit over $N\le\num{18}$ gives $\alpha=\num{2.216}\pm\num{0.034}$, growth slows beyond $N\approx\num{21}$ and reaches $\approx\num{350}$ at $N=\num{64}$ (\cref{SI-sec:si_finite_size}). (b) Static QFI versus $\eta$ for $N=\num{4},\num{8},\num{16},\num{32}$; the thermodynamic critical coupling $\eta_c=\num{0.5}$ is marked.}
    \label{fig:dicke_scaling}
\end{figure}

Spreading the dressing over $N$ qubits lifts the single-qubit ceiling set by two-level saturation. In the superradiant regime of the finite Dicke calculation ($\eta=\num{1.15}$, $\omega_q/\omega_c=\num{1.005}$), the static QFI grows as $N^{\alpha}$ with $\alpha=\num{2.216}\pm\num{0.034}$ (\qty{95}{\percent} interval \numrange{2.150}{2.282}, $R^2=\num{0.9988}$) over $N\le\num{18}$ (\cref{fig:dicke_scaling}a, \cref{tab:trackB_dicke})~\cite{Emary2003,Hotter2025}. This small-$N$ exponent exceeds the linear scaling of independent qubits. Growth then slows: the local exponent drops below unity beyond $N\approx\num{21}$, the QFI moves from \num{229} at $N=\num{27}$ to \num{238} at $N=\num{33}$, and the extended grid reaches $\approx\num{350}$ at $N=\num{64}$ without settling on any single-plateau form (\cref{SI-sec:si_finite_size}). The super-linear exponent therefore describes the small-$N$ window and is not an asymptotic scaling law.

\begin{table}[hbt!]
\centering
\caption{Static Dicke QFI at the fixed operating point $\eta=\num{1.15}$ relative to the coherent baseline. The $N=\num{4}$, \num{8} and \num{16} entries belong to the $\omega_q/\omega_c=\num{1.005}$ family that carries the scaling fit above; the $N=\num{32}$ entry is the $\omega_q/\omega_c=\num{1.0}$ reference (\num{228.6}) used to normalize the collective DCE gain. The QFI maximum over $\eta$ shifts with $N$ (to $\eta\approx\num{1.40}$ at $N=\num{4}$, \num{1.15} at $N=\num{16}$, \num{1.10} at $N=\num{32}$).}
\label{tab:trackB_dicke}
\begin{tabular}{c c}
\toprule
$N$ & $F_Q/F_Q^{\rm coh}$ \\
\midrule
\num{4} & \num{6.4} \\
\num{8} & \num{28.4} \\
\num{16} & \num{134.2} \\
\num{32} & \num{229} \\
\bottomrule
\end{tabular}
\end{table}

The finite-size spectrum at $\eta=\num{1.15}$ keeps the parity structure of the thermodynamic transition, with two nearly degenerate, macroscopically distinct minima~\cite{Emary2003,Kirton2019}. The pure-state QFI presumes that a symmetry-breaking preparation selects one of them; thermal mixing of the two would reduce it. Like the single-qubit gain, this collective gain is non-emissive, and reading it again requires the drive.

\subsection{The collective drive meets the same barrier}
\label{sec:dicke_dce}

\begin{figure}[tbp]
    \centering
    \includegraphics[width=.98\columnwidth]{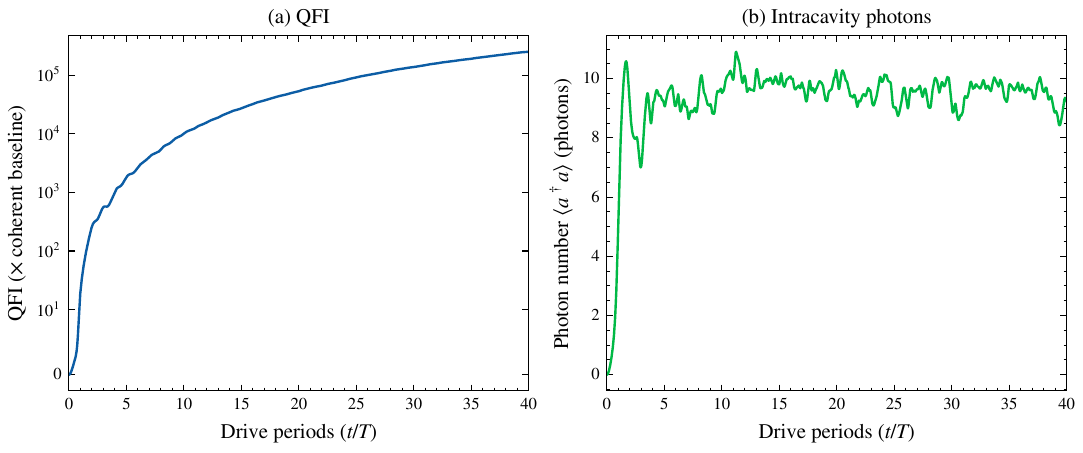}
    \caption{The collective drive builds a plateau of real photons. Dicke+DCE dynamics at $N=\num{32}$, $\eta=\num{1.15}$, $g_d/g_0=\num{0.15}$. (a) The QFI builds over $\approx\num{25}$ drive periods to a late-window average of \num{167271} (\num{730} times the static value) and still rises at the last stored point, \num{251557} at period \num{40} (\cref{SI-sec:si_dicke_convergence}). (b) The intracavity population grows from the vacuum to $n_{\rm peak}\approx\num{10.9}$, with mean $n_{\rm avg}\approx\num{9.3}$ over the window.}
    \label{fig:dicke_dce_oscillations}
\end{figure}

Driving the $N=\num{32}$ Dicke probe at its DCE resonance multiplies the collective gain by \num{730} and then exposes it to the same port. At $g_d/g_0=\num{0.15}$, $\omega_q/\omega_c=\num{1.0}$ and $\omega_d/\omega_c=\num{2.0}$, the late-window average of the pure-state QFI over \num{40} drive periods is \num{167271} at $N_{\rm cav}=\num{28}$ (\cref{tab:trackB_dicke_dce}, \cref{fig:dicke_dce_oscillations}). The collective probe reaches this plateau after $\approx\num{25}$ drive periods rather than oscillating on the \num{34}-period single-qubit cycle. Its endpoint keeps rising with the window, by a factor \num{7.0} from period \num{20} to period \num{60} (\cref{SI-sec:si_dicke_convergence}); the late-window average is the operational figure and the rising endpoint (\num{251557} at period \num{40}) is not a target.

\begin{table}[hbt!]
\centering
\caption{Collective DCE gain ($N=\num{32}$, $\eta=\num{1.15}$, $\omega_q/\omega_c=\num{1.0}$, $\omega_d/\omega_c=\num{2.0}$, $N_{\rm cav}=\num{28}$, \num{40} drive periods, pure state). Enhancement is relative to the static value \num{229}.}
\label{tab:trackB_dicke_dce}
\begin{tabular}{c c c}
\toprule
$g_d/g_0$ & Late-window average QFI & Enhancement over static \\
\midrule
\num{0.15} & \num{167271} & \num{730} \\
\bottomrule
\end{tabular}
\end{table}

\paragraph{The physical-point retention is a truncation sequence.} At the CPW linewidth, the retention of the collective probe relative to each basis's own lossless peak falls monotonically from \qty{16}{\percent} at $N_{\rm cav}=\num{6}$ to \qty{2.6}{\percent} at $N_{\rm cav}=\num{24}$ (\cref{fig:extraction_law}b; \cref{SI-tab:si_dicke_loss_convergence}). A tight photon basis inflates the apparent retention, because hard reflection at the Fock boundary traps phase-sensitive population in the low-photon manifold: at $N_{\rm cav}=\num{6}$ ($\braket{n}\approx\num{3.5}$) the $n\le2$ block alone overshoots the full-space peak by a factor \num{2.65} without loss and \num{1.28} at the physical $\kappa$, whereas at $N_{\rm cav}=\num{12}$ the sector contributions grow monotonically to the full value (\cref{SI-tab:si_photon_sector}). Opening the ladder lets the state reach the Fock components where loss acts, and the retention falls. The \qty{2.6}{\percent} value at the largest basis is therefore a finite-basis point, and the converged retention lies at or below it if the trend persists.

\begin{figure}[htbp]
    \centering
    \includegraphics[width=.97\columnwidth]{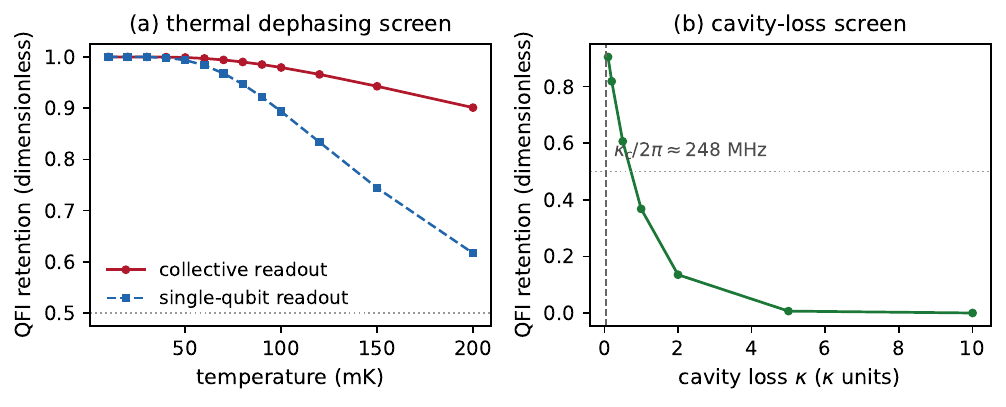}
    \caption{Amplitude-only loss models miss the barrier. (a) QFI retention under the phenomenological thermal-dephasing model for collective and single-qubit readout. (b) Phenomenological amplitude-decay model of cavity loss: it predicts \qty{50}{\percent} degradation near the CPW linewidth (marked at the \qty{248}{\mega\hertz} literature value), whereas the master equation gives \qty{99}{\percent} at $\kappa/2\pi=\qty{246}{\mega\hertz}$, because dissipation reshapes the whole driven evolution.}
    \label{fig:dicke_no_go}
\end{figure}

\paragraph{Cavity loss, not qubit decoherence, dominates.} For the single-qubit probe, qubit relaxation and dephasing at the coherence computed for this circuit ($T_1=\qty{25.8}{\micro\second}$, $T_2=\qty{35.3}{\micro\second}$; \cref{SI-sec:si_qubit_loss}) retain \qty{99.76}{\percent} of the lossless peak, and combining them with the physical cavity loss returns \qty{0.962}{\percent}, indistinguishable from cavity loss alone. The interrogation lasts \num{40} drive periods, \qty{3.2}{\nano\second} at $\omega_c/2\pi=\qty{6}{\giga\hertz}$, so by \cref{eq:eta_eff} the qubit factor $e^{-\gamma_q\tau_{\rm int}}$ costs ten percent only for $T_2\lesssim\qty{30}{\nano\second}$. Cavity loss and collective thermal dephasing also differ in kind: a phenomenological $1/\sqrt{N}$ suppression of collective dephasing (\cref{fig:dicke_no_go}a; \cref{SI-sec:si_thermal}) offers no protection against photon loss, and an amplitude-only decay model underestimates the Lindblad degradation at the physical point by a factor of two, \qty{50}{\percent} against \qty{99}{\percent} (\cref{fig:dicke_no_go}b).

Thermal dephasing and cavity loss do not enter the collective gain on the same footing, and \cref{fig:dicke_no_go} separates them. The assumed $1/\sqrt{N}$ collective-noise suppression gives the displayed retention within this phenomenological model, but it does not include all qubit-qubit interactions, inhomogeneous couplings, drive noise, or a full mixed-state QFI calculation. These curves isolate the collective thermal-dephasing scaling implied by that phenomenological assumption.

\subsection{Operating points, platforms, and a decision protocol}
\label{sec:optimization}

The usable gain of a real device is set by the retained fraction at the drive it can sustain, not by the pure-state ceiling. \Cref{tab:trackB_optimal} pairs each protocol's pure-state optimum with what survives at the CPW linewidth.

\begin{table}[hbtp]
\centering
\caption{Pure-state optima and retained QFI at the CPW operating point $\kappa/2\pi=\qty{246}{\mega\hertz}$ ($\kappa/\omega_c=\num{4.1e-2}$). Single-qubit DCE: maximum of the fixed-detuning scan. Dicke+DCE: \num{40}-period late-window average at $N_{\rm cav}=\num{28}$; its retention is evaluated over \num{10} periods for $N_{\rm cav}=\numrange{6}{24}$ relative to each basis's lossless reference (\cref{SI-sec:si_dicke_loss}).}
\label{tab:trackB_optimal}
\small
\setlength{\tabcolsep}{4pt}
\begin{tabular}{l c c c c l}
\toprule
Protocol & $N$ & $\eta$ & $g_d/g_0$ & Pure-state QFI & Retained at CPW linewidth \\
\midrule
Static Rabi & \num{1} & \num{1.14} & \num{0} & \num{6.12} & non-emissive \\
DCE Rabi & \num{1} & \num{0.150} & \num{0.11923} & \num{1.7e5} & \qty{0.96}{\percent} (QFI \num{1642}) \\
Static Dicke & \num{32} & \num{1.15} & \num{0} & \num{229} & non-emissive \\
Dicke+DCE & \num{32} & \num{1.15} & \num{0.15} & \num{1.7e5} & \qty{16}{\percent} ($N_{\rm cav}=\num{6}$) to \qty{2.6}{\percent} ($N_{\rm cav}=\num{24}$) \\
\bottomrule
\end{tabular}
\end{table}

\paragraph{Multilevel leakage lowers the usable drive.} Real emitters, superconducting qubits and spin ensembles alike~\cite{YouNori2011,Kockum2014,Bertet2014Expe,Putz2017Circ}, have a third level, and the reference drive $g_d/g_0=\num{0.11923}$ pushes population into it. At the physical loss, the \qty{1}{\percent} leakage ceiling admits drives up to $g_d/g_0=\num{0.02}$ for a transmon-like spectrum, \num{0.04} for a flux-qubit-like spectrum and \num{0.06} for a fluxonium-like spectrum, with QFI gains of \num{54.2}, \num{77.5} and \num{85.5} at those drives (\cref{tab:trackB_platforms}; \cref{SI-sec:si_multilevel}). The $1.7\times10^{5}$ peak is thus the two-level ceiling of the calculation, and a device that respects its own anharmonicity operates in the $54$--$86$ window, where the extraction law, not the pure-state ceiling, governs the retained gain.

\begin{table}[hbt!]
\centering
\caption{Platform benchmark. Safe $g_d/g_0$: largest tested drive keeping leakage outside the qubit subspace below \qty{1}{\percent} over \num{40} drive periods at $\kappa/\omega_c=\num{4.1e-2}$ (three-level spectra). The linewidths are representative literature values; every computed entry uses $\kappa/2\pi=\qty{246}{\mega\hertz}$ unless stated.}
\label{tab:trackB_platforms}
\small
\begin{tabular}{l c c c c}
\toprule
Platform & Max.\ $\eta$ & $\kappa/2\pi$ & Safe $g_d/g_0$ & QFI gain \\
\midrule
CPW + flux qubit~\cite{Niemczyk2010,Yoshihara2017} & up to \num{1.3} & $\approx\qty{248}{\mega\hertz}$ & \num{0.04} & \num{77.5} \\
3D cavity~\cite{Paik2011} & --- & $\approx\qty{0.1}{\mega\hertz}$ & --- & DCE retention \qtyrange{97}{99}{\percent}$^{a}$ \\
Transmon array~\cite{Koch2007} & tunable coupler & $\sim\qty{1}{\mega\hertz}$ & \num{0.02} & \num{54.2} \\
Fluxonium~\cite{Manucharyan2009,Somoroff2023} & high anharm. & device-dep. & \num{0.06} & \num{85.5} \\
\bottomrule
\multicolumn{5}{l}{\footnotesize $^{a}$Bracketed by sampled rows: \qty{99.1}{\percent} at \qty{0.06}{\mega\hertz}, \qty{97.0}{\percent} at \qty{0.3}{\mega\hertz}.}
\end{tabular}
\end{table}

\paragraph{Decision protocol.} No demonstrated platform yet combines the three ingredients of the collective reference point: $\eta=\num{1.15}$, a linewidth $\kappa/2\pi\lesssim\qty{1}{\mega\hertz}$, and uniform coupling of $N=\num{32}$ qubits. The CPW circuits that reach deep-strong coupling sit far beyond the knee, at $x\gg x_c$, whereas the cited low-loss 3D-cavity platform reaches its sub-megahertz linewidth at dispersive coupling~\cite{Paik2011}. \Cref{tab:design_screen} condenses the results into a four-step protocol that a designer can apply before building a driven USC sensor. A design that clears all four steps operates where the extraction law, not the pure-state ceiling, sets the gain.

\begin{table}[hbtp]
\centering
\caption{Four-step decision protocol for driven ultrastrong-coupling sensors. Each step is a pass/fail test against a result of this work.}
\label{tab:design_screen}
\small
\begin{tabular}{c l l}
\toprule
Step & \parbox[t]{3.4cm}{\raggedright Check} & \parbox[t]{9.6cm}{\raggedright Criterion and consequence} \\
\midrule
1 & \parbox[t]{3.4cm}{\raggedright Accumulated loss $x=\kappa\tau_{\rm int}$} & \parbox[t]{9.6cm}{\raggedright Evaluate $x$ at the interrogation time the protocol needs. At $x=x_c\approx\num{0.13}$ half the information is lost, and beyond it retention falls as $x^{-1.4}$; at the CPW linewidth retention is below \qty{1}{\percent}.\strut} \\
2 & \parbox[t]{3.4cm}{\raggedright Emitter anharmonicity vs.\ drive} & \parbox[t]{9.6cm}{\raggedright Keep $g_d/g_0$ below the platform's leakage ceiling, \numrange{0.02}{0.06} for the spectra examined, well under the reference drive \num{0.11923} of the pure-state peak.\strut} \\
3 & \parbox[t]{3.4cm}{\raggedright Qubit coherence vs.\ window} & \parbox[t]{9.6cm}{\raggedright Compare $\gamma_q\tau_{\rm int}$ with $\kappa\tau_{\rm int}$. For $T_2=\qty{35.3}{\micro\second}$ the qubit factor is negligible beside cavity decay; cavity loss is the term to reduce first.\strut} \\
4 & \parbox[t]{3.4cm}{\raggedright Nuisance parameters} & \parbox[t]{9.6cm}{\raggedright If $\omega_q$ drifts or is uncalibrated, interrogate through the DCE resonance: the static Fisher matrix is collinear ($\rho_{\eta\omega}=\num{-0.998}$), the driven one keeps \qtyrange{23}{50}{\percent} of the coupling information.\strut} \\
\bottomrule
\end{tabular}
\end{table}

\section{Discussion}
\label{sec:discussion}

Making the virtual photons of an ultrastrongly coupled probe real exposes them to the cavity port that removes them, so the retained metrological gain follows the accumulated cavity loss rather than the pure-state ceiling. The results above make this statement quantitative at three levels. For any loss applied after encoding, monotonicity of the QFI guarantees $R_Q\le1$ without reference to a model. For the driven USC probe, every computed configuration lies not only below unity but below the bare transmissivity of the port. And across both probe families, the size of the deficit is captured by one empirical law in the accumulated loss.

\subsection{The barrier is a conversion cost}
\label{sec:interpretation_B}

The weaker reading of an extraction barrier is that the cavity linewidth attenuates a signal that the drive has already built, and the data exclude it. If the port merely attenuated the output, retention would track $e^{-\kappa\tau_{\rm int}}$; instead it falls a factor \num{24} below that envelope at the CPW linewidth. The drive builds the parameter-sensitive state out of the very photons that the port drains, so loss during the drive changes which state is built, not only how much of it escapes. Static and driven ceilings are therefore bounds of different kinds. The static ceiling is the sensitivity of a ground state that never radiates. The driven ceiling is the sensitivity of a state whose metrological content is exactly what the port can carry away, and it is attainable only by creating the photons that the port removes. Floquet protection of emitter-local baths~\cite{Bai2023} does not change this conclusion, because the cavity leakage channel is broadband and every carrier of the signal crosses it.

The same reasoning explains why the barrier is expressed in the accumulated loss rather than in the linewidth or the time separately. Loss shortens the useful interrogation in proportion to its strength (\cref{sec:extraction}), so a faster readout does not escape the barrier; it moves the operating point along the same curve. Driving harder does not escape it either: beyond the resonance the pump-normalized information falls by three orders of magnitude (\cref{sec:dce}).

\subsection{How the barrier delimits prior claims}
\label{sec:prior_work}

The extraction barrier and the neighboring results answer different questions, and placing them on one axis---estimand, extraction channel, loss budget---shows where each applies (\cref{tab:comparison}).

\begin{table}[t]
\caption{Four accounts of metrological gain from dressed or driven light--matter systems, ordered by how the information leaves the system.}
\label{tab:comparison}
\footnotesize
\setlength{\tabcolsep}{3pt}
\begin{ruledtabular}
\begin{tabular}{l l l l l}
\parbox[t]{2.3cm}{\raggedright Study} & \parbox[t]{2.2cm}{\raggedright Estimand} & \parbox[t]{2.4cm}{\raggedright Extraction} & \parbox[t]{2.4cm}{\raggedright Loss budget} & \parbox[t]{4.3cm}{\raggedright Conclusion within its setting} \\
\colrule
\parbox[t]{2.3cm}{\raggedright Hotter \emph{et al.}~\cite{Hotter2025}} & \parbox[t]{2.2cm}{\raggedright mode frequencies} & \parbox[t]{2.4cm}{\raggedright none (internal)} & \parbox[t]{2.4cm}{\raggedright undriven, no port} & \parbox[t]{4.3cm}{\raggedright virtual dressing is metrologically usable} \\
\parbox[t]{2.3cm}{\raggedright Schelb \emph{et al.}~\cite{Schelb2026}} & \parbox[t]{2.2cm}{\raggedright phase, displacement} & \parbox[t]{2.4cm}{\raggedright closed encoding} & \parbox[t]{2.4cm}{\raggedright $\kappa\tau\ll1$} & \parbox[t]{4.3cm}{\raggedright advantage over classical states at matched photon number} \\
\parbox[t]{2.3cm}{\raggedright Wysocki \emph{et al.}~\cite{Wysocki2026}} & \parbox[t]{2.2cm}{\raggedright encoded parameter} & \parbox[t]{2.4cm}{\raggedright erasure of particles} & \parbox[t]{2.4cm}{\raggedright no port, no cavity} & \parbox[t]{4.3cm}{\raggedright scrambling protects $F_Q$ for any surviving fraction above $N/2$} \\
\parbox[t]{2.3cm}{\raggedright This work} & \parbox[t]{2.2cm}{\raggedright coupling $g$ via DCE} & \parbox[t]{2.4cm}{\raggedright continuous output port} & \parbox[t]{2.4cm}{\raggedright $\kappa/\omega_c=\num{4.1e-2}$} & \parbox[t]{4.3cm}{\raggedright retention \qty{0.96}{\percent} (single qubit), \qty{2.6}{\percent} (Dicke, largest basis)} \\
\end{tabular}
\end{ruledtabular}
\end{table}

Hotter, Miranowicz and Gietka showed that virtual excitations can be leveraged without being extracted and predicted QFI growth $F_Q\propto e^{4\xi}$ in the Dicke model~\cite{Hotter2025}; error-mitigation schemes pursue the same end by protecting the encoded precision through virtual purification~\cite{Lin2025Virtual}. Our static calculation agrees with that trend over the finite sizes computed: $F_Q=\num{229}$ at $N=\num{32}$, $\eta=\num{1.15}$ corresponds to an effective squeezing $\xi_{\rm eff}=\tfrac14\ln F_Q=\num{1.36}$. The drive adds a factor \num{730} that their undriven setting does not contain, and the port then removes most of it. The two results bracket the problem: virtual-state sensing locates where the gain is created, and the barrier locates where it is lost. The same complementarity holds for critical metrology, whose gain also resides in the ground-state curvature near a transition~\cite{Garbe2020,Chu2021,Hotter2024}.

Schelb, Costa and Dodonov report a single-qubit DCE metrological advantage that persists with dissipation~\cite{Schelb2026}. Their estimand is a phase or displacement of the generated field, benchmarked against classical states with the same mean photon number; ours is the coupling $g$ itself, followed through the open port to a fixed linewidth. A closed encoding at short dwell time samples the cavity over a time short compared with $1/\kappa$, which places it at small $x$ on the curve of \cref{fig:extraction_law}a, where retention is near unity. The two results are therefore consistent, and \cref{eq:eta_eff,eq:extraction_law} quantify how far a protocol can lengthen its dwell time before the port erases the advantage they report.

Wysocki, Chwede\'nczuk and P{\l}odzie\'n showed that after a scrambling unitary any surviving subsystem larger than half the system retains the full QFI, because scrambling disperses the parameter information away from the fragile low-rank coherence that erasure would destroy~\cite{Wysocki2026}. That result protects information that remains \emph{inside} the sample after part of it is erased. The extraction barrier concerns information carried \emph{out} of the sample: every quantum that carries the parameter must traverse the port, and \cref{eq:eta_eff} applies to each carrier. Scrambling protects an architecture against the deletion of qubits or modes, not against the finite transmissivity of its readout. Our two companion analyses reach the same conclusion from other directions: a protected Floquet invariant does not by itself protect the nonclassical resource it labels~\cite{Deuyekbe2026ResourceInvariant}, and in non-Hermitian bosonic ladders the forward gain alone fixes the share of the Fisher information carried by the mean response, whatever the lattice architecture~\cite{Deuyekbe2026Petermann}.

\subsection{Structural escapes}
\label{sec:roadmap}

Splitting the readout across $M$ ports of one cavity does not evade the barrier. Each port collects the fraction $\kappa_{\rm ext}/\kappa$ with $\kappa_{\rm ext}=M\kappa_p$, independent of $M$; the master equation with one, two and four ports reproduces the single-port retention exactly, \num{0.0096} at the physical loss against an interrogation-window decay factor of \num{0.227}. A lattice of cavities~\cite{Seok2013Mult} offers no escape for the reason given by the companion ladder analysis~\cite{Deuyekbe2026Petermann}. The remaining levers are the ones the extraction law names: a lower linewidth, a shorter interrogation at fixed information rate, and drives that respect the leakage ceiling. An experimental program follows directly. At $N=1$ the first milestone is to measure the retention curve of \cref{sec:dce} in a low-loss cavity at $\omega_c/2\pi=\qty{6}{\giga\hertz}$ and compare it with the computed curve within a \qty{20}{\percent} calibration margin. At $N=\numrange{4}{8}$ it is to confirm the super-linear static exponent in the loss-controlled regime with the drive off. At $N=\numrange{16}{32}$ collective dressing, DCE modulation and the loss barrier must be demonstrated together, against the loss curve at matched linewidth, truncation and window.

\subsection{Scope and model sensitivity}
\label{sec:limitations_B}

The results hold for the finite-dimensional Rabi and Dicke models, the fixed-detuning drive, two-level emitters below their leakage ceilings, and Markovian cavity decay; \cref{SI-tab:si_scope_matrix} sets each assumption against its quantitative boundary. Four of these assumptions bear directly on the numbers. (i) Memory of the cavity bath: at the physical linewidth the structured-bath HEOM states differ from the Lindblad states by a mean infidelity of \qtyrange{4.8}{24}{\percent}, growing with the Drude cutoff and the observation window, in the \num{18} of \num{48} configurations at this linewidth whose hierarchy remains positive (\cref{SI-sec:si_cavity_heom}); the Lindblad retentions are first-order estimates within this model-comparison range, which is not a statistical error bar. (ii) Truncation: the collective retention is a decreasing sequence in $N_{\rm cav}$, not a converged value. (iii) Representation: the effective modulated coupling is not transformed together with the diamagnetic and control terms, so gauge-consistent predictions for strongly nonadiabatic modulation~\cite{Stokes2021,DiStefano2019} and dressed-basis dissipators~\cite{Beaudoin2011,Settineri2018} remain to be included. (iv) Screens: the thermal and collective-dephasing panels use phenomenological models, the flux-noise amplitude \num{5e-7}~$\Phi_0/\sqrt{\unit{\hertz}}$ enters only as a static input, and the Dicke pure-state QFI presumes a symmetry-breaking preparation whose cost is not modeled. The extraction law is an empirical summary of eighteen configurations; it has been ranked against three simple alternatives and cross-validated by refitting, and it separates into family-specific knees rather than a universal one.

\section{Conclusion}
\label{sec:conclusion}

The metrological gain of ultrastrong coupling is real, but it is carried by the very photons that the port removes. A static dressed vacuum holds its gain where no detector can reach it; a dynamical-Casimir drive makes the gain readable by turning it into radiation, and the radiation leaves through the same port that was meant to deliver it. At the linewidth of coplanar-waveguide circuits the port erases more than \qty{99}{\percent} of the single-qubit gain, and the collective gain fares no better once the photon basis is opened. Because both families collapse onto one law of the accumulated loss, the extraction cost can be read off before a device is built: half the information is gone at $\kappa\tau_{\rm int}\approx\num{0.13}$.

The consequence for USC sensing is a change of figure of merit. A proposal that converts bound vacuum correlations into radiating excitations should be judged by the information that survives its port at the drive its emitters tolerate, not by its lossless ceiling. The same accounting applies to phononic, magnonic and hybrid probes whose metrological resource lives in bound correlations and must cross a lossy interface to be measured. Extraction is the step that makes the gain usable, and it is the step that erases it.

\section*{Supplemental Material}

The Supplemental Material~\cite{TrackB_SM} gives the model definitions and QFI conventions, numerical convergence diagnostics, the single-qubit and collective cavity-loss calculations, the cavity-channel HEOM cross-check, the multilevel-emitter leakage screen, finite-size Dicke scaling, experimental feasibility, the scope matrix, and the extraction-law fit.

\section*{Data availability}

All data supporting the findings of this study are available within the article and its Supplementary material. Manuscript sources, simulation scripts, and the solver records behind every figure and table---the single-qubit Floquet scans, the cavity-memory benchmarks, and the reduced-basis mixed-state loss matrix---accompany the submission and will be deposited in a persistent public archive with a registered identifier upon acceptance.

\section*{Acknowledgments}

We thank the Stellenbosch Institute for Advanced Study for hospitality. Computational resources were provided by the penavoraserver cluster (ThinkStation P720, 128 GB RAM, 48 cores) through Penabei Samafou. This work received no external funding.

\section*{Author contributions}

JD and PD performed the numerical simulations and analyzed the data. NM contributed to the theoretical framework and provided critical feedback on the results. SGNE conceived and supervised the project and directed the research. PD and SGNE wrote the final version of the manuscript. All authors reviewed and approved the final manuscript.

\section*{Competing interests}

The authors declare no competing interests.

\section*{Use of AI}

AI-based language tools assisted with literature searches, script debugging, reference verification and language editing. The authors verified all output and take full responsibility for the manuscript.

\bibliographystyle{apsrev4-2}
\bibliography{04_track_B_metrology_references}

@article{Caves1981,
  title = {Quantum-mechanical noise in an interferometer},
  author = {Caves, Carlton M.},
  journal = {Phys. Rev. D},
  volume = {23},
  issue = {8},
  pages = {1693--1708},
  year = {1981},
  publisher = {American Physical Society},
  doi = {10.1103/PhysRevD.23.1693}
}

@article{Ciuti2005,
  title = {Quantum vacuum properties of the intersubband cavity polariton field},
  author = {Ciuti, Cristiano and Bastard, G\'erald and Carusotto, Iacopo},
  journal = {Phys. Rev. B},
  volume = {72},
  issue = {11},
  pages = {115303},
  year = {2005},
  publisher = {American Physical Society},
  doi = {10.1103/PhysRevB.72.115303}
}

@article{DeLiberato2017,
  title = {Virtual photons in the ground state of a dissipative system},
  author = {De Liberato, Simone},
  journal = {Nat. Commun.},
  volume = {8},
  pages = {1465},
  year = {2017},
  publisher = {Nature Publishing Group},
  doi = {10.1038/s41467-017-01504-5}
}

@article{Stassi2013,
  title = {Spontaneous conversion from virtual to real photons in the ultrastrong-coupling regime},
  author = {Stassi, R. and Ridolfo, A. and Di Stefano, O. and Hartmann, M. J. and Savasta, S.},
  journal = {Phys. Rev. Lett.},
  volume = {110},
  issue = {24},
  pages = {243601},
  year = {2013},
  publisher = {American Physical Society},
  doi = {10.1103/PhysRevLett.110.243601}
}

@article{Garziano2013,
  title = {Switching on and off of ultrastrong light-matter interaction: Photon statistics of quantum vacuum radiation},
  author = {Garziano, L. and Ridolfo, A. and Stassi, R. and Di Stefano, O. and Savasta, S.},
  journal = {Phys. Rev. A},
  volume = {88},
  issue = {6},
  pages = {063829},
  year = {2013},
  publisher = {American Physical Society},
  doi = {10.1103/PhysRevA.88.063829}
}

@article{Giannelli2024,
  title = {Detecting virtual photons in ultrastrongly coupled superconducting quantum circuits},
  author = {Giannelli, Luigi and Paladino, Elisabetta and Grajcar, Miroslav and Paraoanu, Gheorghe Sorin and Falci, Giuseppe},
  journal = {Phys. Rev. Research},
  volume = {6},
  issue = {1},
  pages = {013008},
  year = {2024},
  publisher = {American Physical Society},
  doi = {10.1103/PhysRevResearch.6.013008}
}

@article{Stokes2021,
  title = {Ultrastrong time-dependent light-matter interactions are gauge relative},
  author = {Stokes, Adam and Nazir, Ahsan},
  journal = {Phys. Rev. Research},
  volume = {3},
  issue = {1},
  pages = {013116},
  year = {2021},
  publisher = {American Physical Society},
  doi = {10.1103/PhysRevResearch.3.013116}
}

@article{Macri2018,
  title = {Nonperturbative dynamical {C}asimir effect in optomechanical systems: Vacuum {C}asimir--{R}abi splittings},
  author = {Macr\`i, Vincenzo and Ridolfo, Alessandro and Di Stefano, Omar and Kockum, Anton Frisk and Nori, Franco and Savasta, Salvatore},
  journal = {Phys. Rev. X},
  volume = {8},
  issue = {1},
  pages = {011031},
  year = {2018},
  publisher = {American Physical Society},
  doi = {10.1103/PhysRevX.8.011031}
}

@article{Hotter2025,
  title = {Quantum metrology in the ultrastrong coupling regime of light-matter interactions: Leveraging virtual excitations without extracting them},
  author = {Hotter, Christoph and Miranowicz, Adam and Gietka, Karol},
  journal = {Phys. Rev. Lett.},
  volume = {135},
  issue = {10},
  pages = {100802},
  year = {2025},
  publisher = {American Physical Society},
  doi = {10.1103/cxvs-5pb1}
}

@article{Wilson2011,
  title = {Observation of the dynamical Casimir effect in a superconducting circuit},
  author = {Wilson, C. M. and Johansson, G. and Pourkabirian, A. and Simoen, M. and Johansson, J. R. and Duty, T. and Nori, F. and Delsing, P.},
  journal = {Nature},
  volume = {479},
  issue = {7373},
  pages = {376--379},
  year = {2011},
  publisher = {Nature Publishing Group},
  doi = {10.1038/nature10561}
}

@article{Pezze2018,
  title = {Quantum metrology with nonclassical states of atomic ensembles},
  author = {Pezz\`e, Luca and Smerzi, Augusto and Oberthaler, Markus K. and Schmied, Roman and Treutlein, Philipp},
  journal = {Rev. Mod. Phys.},
  volume = {90},
  issue = {3},
  pages = {035005},
  year = {2018},
  publisher = {American Physical Society},
  doi = {10.1103/RevModPhys.90.035005}
}

@article{Giovannetti2004,
  title = {Quantum-enhanced measurements: beating the standard quantum limit},
  author = {Giovannetti, Vittorio and Lloyd, Seth and Maccone, Lorenzo},
  journal = {Science},
  volume = {306},
  issue = {5700},
  pages = {1330--1336},
  year = {2004},
  publisher = {American Association for the Advancement of Science},
  doi = {10.1126/science.1104149}
}

@article{Giovannetti2011,
  title = {Advances in quantum metrology},
  author = {Giovannetti, Vittorio and Lloyd, Seth and Maccone, Lorenzo},
  journal = {Nat. Photon.},
  volume = {5},
  pages = {222--229},
  year = {2011},
  publisher = {Nature Publishing Group},
  doi = {10.1038/nphoton.2011.35}
}

@article{BraunsteinCaves1994,
  title = {Statistical distance and the geometry of quantum states},
  author = {Braunstein, Samuel L. and Caves, Carlton M.},
  journal = {Phys. Rev. Lett.},
  volume = {72},
  issue = {22},
  pages = {3439--3443},
  year = {1994},
  publisher = {American Physical Society},
  doi = {10.1103/PhysRevLett.72.3439}
}

@article{Escher2011,
  title = {General framework for estimating the ultimate precision limit in noisy quantum-enhanced metrology},
  author = {Escher, Bruno M. and de Matos Filho, R. L. and Davidovich, Luiz},
  journal = {Nat. Phys.},
  volume = {7},
  pages = {406--411},
  year = {2011},
  publisher = {Nature Publishing Group},
  doi = {10.1038/nphys1958}
}

@article{Kolodynski2010,
  title = {Phase estimation without a priori phase knowledge in the presence of loss},
  author = {Ko{\l}ody{\'n}ski, Jan and Demkowicz-Dobrza{\'n}ski, Rafa{\l}},
  journal = {Phys. Rev. A},
  volume = {82},
  issue = {5},
  pages = {053804},
  year = {2010},
  publisher = {American Physical Society},
  doi = {10.1103/PhysRevA.82.053804}
}

@article{Demkowicz2012,
  title = {The elusive {H}eisenberg limit in quantum-enhanced metrology},
  author = {Demkowicz-Dobrza{\'n}ski, Rafa{\l} and Ko{\l}ody{\'n}ski, Jan and Gu{\c t}{\u a}, M{\u a}d{\u a}lin},
  journal = {Nat. Commun.},
  volume = {3},
  pages = {1063},
  year = {2012},
  publisher = {Nature Publishing Group},
  doi = {10.1038/ncomms2067}
}

@article{Szczykulska2016,
  title = {Multi-parameter quantum metrology},
  author = {Szczykulska, Magdalena and Baumgratz, Tillmann and Datta, Animesh},
  journal = {Adv. Phys. X},
  volume = {1},
  issue = {4},
  pages = {621--639},
  year = {2016},
  publisher = {Taylor \& Francis},
  doi = {10.1080/23746149.2016.1230476}
}

@article{Ragy2016,
  title = {Compatibility in multiparameter quantum metrology},
  author = {Ragy, Sammy and Jarzyna, Marcin and Demkowicz-Dobrza{\'n}ski, Rafa{\l}},
  journal = {Phys. Rev. A},
  volume = {94},
  issue = {5},
  pages = {052108},
  year = {2016},
  publisher = {American Physical Society},
  doi = {10.1103/PhysRevA.94.052108}
}

@article{Paris2009,
  title = {Quantum estimation for quantum technology},
  author = {Paris, Matteo G. A.},
  journal = {Int. J. Quantum Inf.},
  volume = {7},
  issue = {supp01},
  pages = {125--137},
  year = {2009},
  publisher = {World Scientific},
  doi = {10.1142/S0219749909004839}
}

@article{Clerk2010,
  title = {Introduction to quantum noise, measurement, and amplification},
  author = {Clerk, A. A. and Devoret, M. H. and Girvin, S. M. and Marquardt, Florian and Schoelkopf, R. J.},
  journal = {Rev. Mod. Phys.},
  volume = {82},
  issue = {2},
  pages = {1155--1208},
  year = {2010},
  publisher = {American Physical Society},
  doi = {10.1103/RevModPhys.82.1155}
}

@article{Niemczyk2010,
  title = {Circuit quantum electrodynamics in the ultrastrong-coupling regime},
  author = {Niemczyk, T. and Deppe, F. and Huebl, H. and Menzel, E. P. and Hoi, F. and Marx, A. and Gross, R. and Fink, M. J. and Schwarz, M. J. and Goppl, M. and Romero, G. and Solano, E. and Filipp, S.},
  journal = {Nat. Phys.},
  volume = {6},
  pages = {772--776},
  year = {2010},
  publisher = {Nature Publishing Group},
  doi = {10.1038/nphys1730}
}

@article{FornDiaz2017,
  title = {Ultrastrong coupling of a single artificial atom to an electromagnetic continuum in the nonperturbative regime},
  author = {Forn-D\'iaz, P. and Garc\'ia-Ripoll, J. J. and Peropadre, B. and Orgiazzi, J.-L. and Yurtalan, M. A. and Belyansky, R. and Wilson, C. M. and Lupascu, A.},
  journal = {Nat. Phys.},
  volume = {13},
  pages = {39--43},
  year = {2017},
  publisher = {Nature Publishing Group},
  doi = {10.1038/nphys3905}
}

@article{FornDiaz2019,
  title = {Ultrastrong coupling regimes of light-matter interaction},
  author = {Forn-D\'iaz, P. and Lamata, L. and Rico, E. and Kono, J. and Solano, E.},
  journal = {Rev. Mod. Phys.},
  volume = {91},
  issue = {2},
  pages = {025005},
  year = {2019},
  publisher = {American Physical Society},
  doi = {10.1103/RevModPhys.91.025005}
}

@article{Kockum2019,
  title = {Ultrastrong coupling between light and matter},
  author = {Kockum, Anton Frisk and Miranowicz, Adam and De Liberato, Simone and Savasta, Salvatore and Nori, Franco},
  journal = {Nat. Rev. Phys.},
  volume = {1},
  pages = {19--40},
  year = {2019},
  publisher = {Nature Publishing Group},
  doi = {10.1038/s42254-018-0006-2}
}

@article{Yoshihara2017,
  title = {Superconducting qubit--oscillator circuit beyond the ultrastrong-coupling regime},
  author = {Yoshihara, F. and Furuya, K. and Ashhab, S. and Nori, Franco and Semba, K.},
  journal = {Nat. Phys.},
  volume = {13},
  pages = {44--47},
  year = {2017},
  publisher = {Nature Publishing Group},
  doi = {10.1038/nphys3906}
}

@phdthesis{Kockum2014,
  title = {Quantum optics with artificial atoms},
  author = {Kockum, Anton Frisk},
  school = {Chalmers University of Technology},
  year = {2014}
}

@article{Johansson2012,
  title = {QuTiP: An open-source Python framework for the dynamics of open quantum systems},
  author = {Johansson, J. R. and Nation, P. D. and Nori, Franco},
  journal = {Comput. Phys. Commun.},
  volume = {183},
  issue = {8},
  pages = {1760--1772},
  year = {2012},
  publisher = {Elsevier},
  doi = {10.1016/j.cpc.2012.02.021}
}

@article{Larson2024,
  title = {{F}loquet analysis perspective of driven light--matter interaction models},
  author = {Larson, Jonas},
  journal = {J. Opt. Soc. Am. B},
  volume = {41},
  issue = {8},
  pages = {C235--C246},
  year = {2024},
  publisher = {Optica},
  doi = {10.1364/josab.524005}
}

@article{Nation2012,
  title = {Colloquium: Stimulating uncertainty: Amplifying the quantum vacuum with superconducting circuits},
  author = {Nation, P. D. and Johansson, J. R. and Blencowe, M. P. and Nori, Franco},
  journal = {Rev. Mod. Phys.},
  volume = {84},
  issue = {1},
  pages = {1--24},
  year = {2012},
  publisher = {American Physical Society},
  doi = {10.1103/RevModPhys.84.1}
}

@article{Didier2015,
  title = {Fast Quantum Nondemolition Readout by Parametric Modulation of Longitudinal Qubit-Oscillator Interaction},
  author = {Didier, Nicolas and Bourassa, J\'er\^ome and Blais, Alexandre},
  journal = {Phys. Rev. Lett.},
  volume = {115},
  issue = {20},
  pages = {203601},
  year = {2015},
  publisher = {American Physical Society},
  doi = {10.1103/PhysRevLett.115.203601}
}

@article{YouNori2011,
  title = {Atomic physics and quantum optics using superconducting circuits},
  author = {You, J. Q. and Nori, Franco},
  journal = {Nature},
  volume = {474},
  issue = {7353},
  pages = {589--597},
  year = {2011},
  publisher = {Nature Publishing Group},
  doi = {10.1038/nature10122}
}

@article{Lin2025Virtual,
  title = {Error-mitigated quantum metrology via enhanced virtual purification},
  author = {Lin, Xiaodie and Yuan, Haidong},
  journal = {Communications Physics},
  volume = {8},
  pages = {481},
  year = {2025},
  doi = {10.1038/s42005-025-02383-9}
}

@incollection{Bertet2014Expe,
  title = {Experimental Circuit QED},
  author = {Bertet, Patrice},
  booktitle = {Strong Light-Matter Coupling},
  publisher = {World Scientific},
  year = {2014},
  doi = {10.1142/9789814460354_0003}
}

@book{Putz2017Circ,
  title = {Circuit Cavity QED with Macroscopic Solid-State Spin Ensembles},
  author = {Putz, Stefan},
  publisher = {Springer},
  series = {Springer Theses},
  year = {2017},
  doi = {10.1007/978-3-319-66447-7}
}

@incollection{Girvin2019Schr,
  title = {Schr\"odinger cat states in circuit QED},
  author = {Girvin, Steven M.},
  booktitle = {Current Trends in Atomic Physics},
  publisher = {Oxford University Press},
  year = {2019},
  doi = {10.1093/oso/9780198837190.003.0011}
}

@article{Seok2013Mult,
  title = {Multimode strong-coupling quantum optomechanics},
  author = {Seok, H. and Buchmann, L. F. and Wright, E. M. and Meystre, P.},
  journal = {Phys. Rev. A},
  volume = {88},
  issue = {6},
  pages = {063850},
  year = {2013},
  publisher = {American Physical Society},
  doi = {10.1103/PhysRevA.88.063850}
}

@article{Gilmore1976Pers,
  title = {Persistence of the phase transition in the Dicke model with external fields and counter-rotating terms},
  author = {Gilmore, R.},
  journal = {Phys. Lett. A},
  volume = {55},
  issue = {8},
  pages = {459--461},
  year = {1976},
  doi = {10.1016/0375-9601(76)90220-6}
}

@article{Akbari2025,
  author    = {Akbari, Kamran and Nori, Franco and Hughes, Stephen},
  journal   = {Phys. Rev. Lett.},
  title     = {Floquet engineering the quantum Rabi model in the ultrastrong coupling regime},
  year      = {2025},
  volume    = {134},
  number    = {6},
  pages     = {063602},
  doi       = {10.1103/PhysRevLett.134.063602},
  publisher = {American Physical Society},
}

@Book{Gardiner2004,
  author    = {Gardiner, Crispin and Zoller, Peter},
  publisher = {Springer},
  title     = {Quantum Noise: A Handbook of Markovian and Non-Markovian Quantum Stochastic Methods with Applications to Quantum Optics},
  year      = {2004},
  edition   = {3rd},
  pages     = {1--524},
}

@article{Yan2016,
  title = {The flux qubit revisited to enhance coherence and reproducibility},
  author = {Yan, Fei and Gustavsson, Simon and Kamal, Archana and Birenbaum, Jeffrey and Sears, Adam P. and Hover, David and Gudmundsen, Ted J. and Rosenberg, Danna and Samach, Gabriel and Weber, Steven and Yoder, Jonilyn L. and Orlando, Terry P. and Clarke, John and Kerman, Andrew J. and Oliver, William D.},
  journal = {Nat. Commun.},
  volume = {7},
  pages = {12964},
  year = {2016},
  publisher = {Nature Publishing Group},
  doi = {10.1038/ncomms12964}
}

@article{Schelb2026,
  title = {Metrological power of single-qubit dynamical {C}asimir effect in circuit {QED}},
  author = {Schelb, H. R. and Costa, A. P. and Dodonov, A. V.},
  journal = {Physics Letters A},
  volume = {566},
  pages = {131163},
  year = {2026},
  doi = {10.1016/j.physleta.2025.131163}
}

@article{Bai2023,
  title = {Floquet Engineering to Overcome No-Go Theorem of Noisy Quantum Metrology},
  author = {Bai, Si-Yuan and An, Jun-Hong},
  journal = {Phys. Rev. Lett.},
  volume = {131},
  issue = {5},
  pages = {050801},
  year = {2023},
  publisher = {American Physical Society},
  doi = {10.1103/PhysRevLett.131.050801}
}

@article{Kurdzialek2025,
  title = {Universal Bounds for Quantum Metrology in the Presence of Correlated Noise},
  author = {Kurdzialek, Stanislaw and Albarelli, Francesco and Demkowicz-Dobrzanski, Rafal},
  journal = {Phys. Rev. Lett.},
  volume = {135},
  issue = {13},
  pages = {130801},
  year = {2025},
  publisher = {American Physical Society},
  doi = {10.1103/jy3v-wkcb}
}

@article{Hotter2024,
  title = {Combining Critical and Quantum Metrology},
  author = {Hotter, Christoph and Ritsch, Helmut and Gietka, Karol},
  journal = {Phys. Rev. Lett.},
  volume = {132},
  issue = {6},
  pages = {060801},
  year = {2024},
  publisher = {American Physical Society},
  doi = {10.1103/PhysRevLett.132.060801}
}

@article{Wysocki2026,
  author  = {Wysocki, Piotr and Chwede\'nczuk, Jan and P{\l}odzie\'n, Marcin},
  title   = {Volume-law protection of metrological advantage},
  journal = {Phys. Rev. Lett.},
  volume  = {137},
  pages   = {030801},
  year    = {2026},
  doi     = {10.1103/3gg8-wgdr}
}

@misc{Deuyekbe2026ResourceInvariant,
  author  = {Deuyekbe, Juste and Djorw\'e, P. and {Nsangou Mama} and Nana Engo, S.~G.},
  title   = {The resource--invariant gap: a protected {F}loquet invariant does not
             protect {W}igner negativity},
  year    = {2026},
  note    = {Submitted to PRX Quantum}
}

@misc{Deuyekbe2026Petermann,
  author  = {Deuyekbe, Juste and Djorw\'e, P. and {Nsangou Mama} and Nana Engo, S.~G.},
  title   = {Spatial channel separation lifts the single-port {P}etermann penalty
             in a non-{H}ermitian optomechanical ladder},
  year    = {2026},
  note    = {Submitted to npj Quantum Information}
}

@misc{TrackB_SM,
  title = {Supplemental Material for ``Extraction erases the gain: cavity loss bounds dynamical-{C}asimir metrology at ultrastrong coupling''},
  year  = {2026},
  note  = {See Supplemental Material at [URL will be inserted by publisher] for model definitions and quantum Fisher information conventions, numerical convergence diagnostics, single-qubit and collective cavity-loss calculations, the cavity-channel hierarchical equations-of-motion cross-check, the multilevel-emitter leakage screen, finite-size Dicke scaling, experimental feasibility assessments, and scope boundaries.}
}

@article{Beaudoin2011,
  title = {Dissipation and ultrastrong coupling in circuit {QED}},
  author = {Beaudoin, F\'elix and Gambetta, Jay M. and Blais, Alexandre},
  journal = {Phys. Rev. A},
  volume = {84},
  pages = {043832},
  year = {2011},
  doi = {10.1103/PhysRevA.84.043832}
}

@article{Settineri2018,
  title = {Dissipation and thermal noise in hybrid quantum systems in the ultrastrong-coupling regime},
  author = {Settineri, Alessio and Macr\`i, Vincenzo and Ridolfo, Alessandro and Di Stefano, Omar and Kockum, Anton Frisk and Nori, Franco and Savasta, Salvatore},
  journal = {Phys. Rev. A},
  volume = {98},
  pages = {053834},
  year = {2018},
  doi = {10.1103/PhysRevA.98.053834}
}

@article{DiStefano2019,
  title = {Resolution of gauge ambiguities in ultrastrong-coupling cavity quantum electrodynamics},
  author = {Di Stefano, Omar and Settineri, Alessio and Macr\`i, Vincenzo and Garziano, Luigi and Stassi, Roberto and Savasta, Salvatore and Nori, Franco},
  journal = {Nat. Phys.},
  volume = {15},
  pages = {803--808},
  year = {2019},
  doi = {10.1038/s41567-019-0534-4}
}

@article{Tanimura2020,
  title = {Numerically ``exact'' approach to open quantum dynamics: The hierarchical equations of motion ({HEOM})},
  author = {Tanimura, Yoshitaka},
  journal = {J. Chem. Phys.},
  volume = {153},
  pages = {020901},
  year = {2020},
  doi = {10.1063/5.0011599}
}

@article{Lambert2023,
  title = {{QuTiP-BoFiN}: A bosonic and fermionic numerical hierarchical-equations-of-motion library with applications in light-harvesting, quantum control, and single-molecule electronics},
  author = {Lambert, Neill and Raheja, Tarun and Cross, Simon and Menczel, Paul and Ahmed, Shahnawaz and Pitchford, Alexander and Burgarth, Daniel and Nori, Franco},
  journal = {Phys. Rev. Res.},
  volume = {5},
  pages = {013181},
  year = {2023},
  doi = {10.1103/PhysRevResearch.5.013181}
}

@article{Liu2020,
  title = {Quantum {F}isher information matrix and multiparameter estimation},
  author = {Liu, Jing and Yuan, Haidong and Lu, Xiao-Ming and Wang, Xiaoguang},
  journal = {J. Phys. A: Math. Theor.},
  volume = {53},
  pages = {023001},
  year = {2020},
  doi = {10.1088/1751-8121/ab5d4d}
}

@article{Dodonov2020,
  title = {Fifty years of the dynamical {C}asimir effect},
  author = {Dodonov, Viktor},
  journal = {Physics},
  volume = {2},
  pages = {67--104},
  year = {2020},
  doi = {10.3390/physics2010007}
}

@article{Johansson2009,
  title = {Dynamical {C}asimir effect in a superconducting coplanar waveguide},
  author = {Johansson, J. R. and Johansson, G. and Wilson, C. M. and Nori, Franco},
  journal = {Phys. Rev. Lett.},
  volume = {103},
  pages = {147003},
  year = {2009},
  doi = {10.1103/PhysRevLett.103.147003}
}

@article{Lahteenmaki2013,
  title = {Dynamical {C}asimir effect in a {J}osephson metamaterial},
  author = {L\"ahteenm\"aki, Pasi and Paraoanu, G. S. and Hassel, Juha and Hakonen, Pertti J.},
  journal = {Proc. Natl. Acad. Sci. USA},
  volume = {110},
  pages = {4234--4238},
  year = {2013},
  doi = {10.1073/pnas.1212705110}
}

@article{Emary2003,
  title = {Chaos and the quantum phase transition in the {D}icke model},
  author = {Emary, Clive and Brandes, Tobias},
  journal = {Phys. Rev. E},
  volume = {67},
  pages = {066203},
  year = {2003},
  doi = {10.1103/PhysRevE.67.066203}
}

@article{Kirton2019,
  title = {Introduction to the {D}icke model: From equilibrium to nonequilibrium, and vice versa},
  author = {Kirton, Peter and Roses, Mor M. and Keeling, Jonathan and Dalla Torre, Emanuele G.},
  journal = {Adv. Quantum Technol.},
  volume = {2},
  pages = {1800043},
  year = {2019},
  doi = {10.1002/qute.201800043}
}

@article{Garbe2020,
  title = {Critical quantum metrology with a finite-component quantum phase transition},
  author = {Garbe, Louis and Bina, Matteo and Keller, Arne and Paris, Matteo G. A. and Felicetti, Simone},
  journal = {Phys. Rev. Lett.},
  volume = {124},
  pages = {120504},
  year = {2020},
  doi = {10.1103/PhysRevLett.124.120504}
}

@article{Chu2021,
  title = {Dynamic framework for criticality-enhanced quantum sensing},
  author = {Chu, Yaoming and Zhang, Shaoliang and Yu, Baiyi and Cai, Jianming},
  journal = {Phys. Rev. Lett.},
  volume = {126},
  pages = {010502},
  year = {2021},
  doi = {10.1103/PhysRevLett.126.010502}
}

@article{Koch2007,
  title = {Charge-insensitive qubit design derived from the {C}ooper pair box},
  author = {Koch, Jens and Yu, Terri M. and Gambetta, Jay and Houck, A. A. and Schuster, D. I. and Majer, J. and Blais, Alexandre and Devoret, M. H. and Girvin, S. M. and Schoelkopf, R. J.},
  journal = {Phys. Rev. A},
  volume = {76},
  pages = {042319},
  year = {2007},
  doi = {10.1103/PhysRevA.76.042319}
}

@article{Manucharyan2009,
  title = {Fluxonium: Single {C}ooper-pair circuit free of charge offsets},
  author = {Manucharyan, Vladimir E. and Koch, Jens and Glazman, Leonid I. and Devoret, Michel H.},
  journal = {Science},
  volume = {326},
  pages = {113--116},
  year = {2009},
  doi = {10.1126/science.1175552}
}

@article{Somoroff2023,
  title = {Millisecond coherence in a superconducting qubit},
  author = {Somoroff, Aaron and Ficheux, Quentin and Mencia, Raymond A. and Xiong, Haonan and Kuzmin, Roman and Manucharyan, Vladimir E.},
  journal = {Phys. Rev. Lett.},
  volume = {130},
  pages = {267001},
  year = {2023},
  doi = {10.1103/PhysRevLett.130.267001}
}

@article{GardinerCollett1985,
  title = {Input and output in damped quantum systems: Quantum stochastic differential equations and the master equation},
  author = {Gardiner, C. W. and Collett, M. J.},
  journal = {Phys. Rev. A},
  volume = {31},
  pages = {3761--3774},
  year = {1985},
  doi = {10.1103/PhysRevA.31.3761}
}

@article{Ashhab2010,
  title = {Qubit-oscillator systems in the ultrastrong-coupling regime and their potential for preparing nonclassical states},
  author = {Ashhab, S. and Nori, Franco},
  journal = {Phys. Rev. A},
  volume = {81},
  pages = {042311},
  year = {2010},
  doi = {10.1103/PhysRevA.81.042311}
}

@article{Zhou2021,
  title = {Asymptotic theory of quantum channel estimation},
  author = {Zhou, Sisi and Jiang, Liang},
  journal = {PRX Quantum},
  volume = {2},
  pages = {010343},
  year = {2021},
  doi = {10.1103/PRXQuantum.2.010343}
}

@article{Paik2011,
  title = {Observation of high coherence in {J}osephson junction qubits measured in a three-dimensional circuit {QED} architecture},
  author = {Paik, Hanhee and Schuster, D. I. and Bishop, Lev S. and Kirchmair, G. and Catelani, G. and Sears, A. P. and Johnson, B. R. and Reagor, M. J. and Frunzio, L. and Glazman, L. I. and Girvin, S. M. and Devoret, M. H. and Schoelkopf, R. J.},
  journal = {Phys. Rev. Lett.},
  volume = {107},
  pages = {240501},
  year = {2011},
  doi = {10.1103/PhysRevLett.107.240501}
}

@unpublished{Deuyekbe2026CompanionGap,
  author = {Deuyekbe, Juste and Djorw\'e, P. and {Nsangou Mama} and Nana Engo, S.~G.},
  title  = {The resource--invariant gap: a protected {F}loquet invariant need not protect
            {W}igner negativity},
  year   = {2026},
  note   = {Companion manuscript, submitted}
}

@unpublished{Deuyekbe2026CompanionLadder,
  author = {Deuyekbe, Juste and Djorw\'e, P. and {Nsangou Mama} and Nana Engo, S.~G.},
  title  = {Forward gain tracks the mean-response share of the {F}isher information in
            non-{H}ermitian bosonic ladders},
  year   = {2026},
  note   = {Companion manuscript, submitted}
}

\end{document}


\title{Supplemental Material:\\Extraction erases the gain: cavity loss bounds dynamical-Casimir metrology at ultrastrong coupling}

\author{Juste Deuyekbe}
\email{justedeuyekbe@gmail.com}
\affiliation{Department of Physics, Faculty of Science, University of Ngaoundere, P.O. Box 454, Ngaoundere, Cameroon}

\author{P. Djorw\'e}
\email{djorwepp@gmail.com}
\affiliation{Department of Physics, Faculty of Science, University of Ngaoundere, P.O. Box 454, Ngaoundere, Cameroon}
\affiliation{Stellenbosch Institute for Advanced Study (STIAS), Stellenbosch 7600, South Africa}

\author{Nsangou Mama}
\email{mnsangou@yahoo.com}
\affiliation{Department of Physics, Faculty of Science, University of Maroua, P.O. Box 46, Maroua, Cameroon}

\author{S. G. Nana Engo}
\email{serge.nana-engo@facsciences-uy1.cm}
\affiliation{Department of Physics, Faculty of Science, University of Yaounde I, P.O. Box 812, Yaounde, Cameroon}

\maketitle

\noindent
This Supplemental Material provides the model definitions and QFI conventions (\cref{sec:si_exact_diag}), numerical convergence diagnostics (\cref{sec:si_convergence}), and the single-qubit and collective loss calculations that produce the main retention values (\cref{sec:si_loss_adjusted,sec:si_dicke_loss}). It also documents the cavity-channel HEOM cross-check (\cref{sec:si_cavity_heom}), the multilevel-emitter leakage screen (\cref{sec:si_multilevel}), the finite-size Dicke scaling (\cref{sec:si_finite_size}), and the additional model-specific screens and experimental feasibility assessments (\cref{sec:si_additional,sec:si_feasibility_B}). The scope boundaries and the empirical extraction fit are given in \cref{sec:si_scope_B,sec:si_extraction}.

\section{Model definitions and QFI conventions}
\label{sec:si_exact_diag}

We set $\hbar=1$ throughout. The single-qubit quantum Rabi Hamiltonian is
\begin{equation}
\hat{H}_{\text{Rabi}} = \frac{\omega_q}{2}\hat{\sigma}_z + \omega_c \hat{a}^\dagger \hat{a} + g \hat{\sigma}_x (\hat{a} + \hat{a}^\dagger),
\label{eq:si_rabi}
\end{equation}
with normalized coupling $\eta = g/\omega_c$. The collective Dicke Hamiltonian for $N$ qubits is
\begin{equation}
\hat{H}_{\text{Dicke}} = \omega_c \hat{a}^\dagger \hat{a} + \omega_q \hat{J}_z + \frac{g}{\sqrt{N}} \hat{J}_x (\hat{a} + \hat{a}^\dagger),
\label{eq:si_dicke}
\end{equation}
with $\hat{J}_{x,z} = \frac{1}{2}\sum_{i=1}^N \hat{\sigma}_{x,z}^{(i)}$. The $1/\sqrt{N}$ normalization gives the thermodynamic critical coupling $\eta_c = \frac{1}{2}\sqrt{\omega_q/\omega_c}$, which equals $\num{0.5}$ at resonance.

Both Hamiltonians are diagonalized in the Fock-qubit (or collective-spin) basis with truncations $n_{\max}=\num{50}$ (Rabi) and $n_{\max}=\num{35}$ (Dicke, $N\le\num{64}$). The driven DCE protocol modulates the coupling as $g(t)=g_0+g_d\cos(\omega_d t)$; the Schr\"odinger equation is integrated with QuTiP~\cite{Johansson2012}. Because the interaction is time dependent, strongly nonadiabatic USC control is gauge relative~\cite{Stokes2021}, so the results below describe the effective bilinear-coupling model; an explicit finite-dimensional similarity transformation verifies internal representation consistency to $\num{5e-9}$ (\cref{sec:si_additional}) without fixing the physical gauge choice~\cite{Stokes2021}.

\subsection{Quantum Fisher information}
\label{sec:si_qfi}

For a static eigenstate family $\hat H(\eta)\ket{\Psi_G(\eta)}=E_0(\eta)\ket{\Psi_G(\eta)}$ with generator $\hat H_{\rm int}=\partial_\eta\hat H$, the exact QFI is the gap-weighted sum
\begin{equation}
F_Q = 4\sum_{k\neq 0}\frac{|\braket{\Psi_k|\hat H_{\text{int}}|\Psi_G}|^2}{(E_k-E_0)^2}.
\label{eq:si_qfi_static}
\end{equation}
The variance form $F_Q=4\,\mathrm{Var}[\hat H_{\rm int}]$ holds only for unitary-encoding families and overestimates the static QFI by a factor $\num{3.4}$ at the sampled maximum ($\num{20.9}$ vs.\ $\num{6.116}$ at $\eta=\num{1.14}$). The reported static values are a fidelity-based finite difference with $\epsilon=\num{1e-6}$; against the exact spectral sum of \cref{eq:si_qfi_static} the two forms agree to better than \qty{0.43}{\percent} across the sweep, the residual being largest at the deepest coupling.

For the driven DCE protocol the state is not stationary, and the QFI is computed from the fidelity
\begin{equation}
F_Q(t) \approx 8\,\frac{1-|\braket{\Psi_\eta(t)|\Psi_{\eta+\delta}(t)}|}{\delta^2},
\qquad \delta=\num{1e-3},
\label{eq:si_qfi_time}
\end{equation}
which becomes exact as $\delta\to0$. All DCE QFI values quoted in the main text are this fidelity-based quantity. At the single-qubit DCE optimum the instantaneous variance form ($\num{4}\,\mathrm{Var}[\hat H_{\rm int}]$) gives \num{225.5}, against the fidelity-based peak of \num{170903}; the two are different quantities and are not interchanged. Convergence with respect to time step and $\delta$ is reported in \cref{sec:si_convergence}.

Two normalizations are used and they are not interchangeable. Ratios against the coherent-state baseline $F_Q^{\rm coh}=1$ are scale markers only: that baseline matches neither the pump power, the intracavity population, the bandwidth, nor the readout cost of any driven comparison, so a number divided by it carries no resource meaning. The convention adopted here separates the two budgets. The encoded resource is $\mathcal{R}_{\rm enc}=\langle \hat n_{\rm cav}\rangle_{\rm ss}$ in quanta, with two further internal resources reported separately and not conflated: the drive-dressing parameter $n_{\rm dr}=g_d^2/\omega_c^2$, which is a drive amplitude squared in cavity units, and the virtual ground-state population $n_{\rm virt}=\braket{\hat a^\dagger\hat a}_{\Psi_G}$, which is what the dressing actually creates ($n_{\rm virt}\approx\eta^2/4$ for the static operator). Added noise, $n_{\rm th}(\omega,T)$ plus the first-stage amplifier contribution, enters only as the denominator of a signal-to-noise ratio. The admission test is that at $\mathcal{R}_{\rm enc}=1$ and with no added noise the convention reproduces the standard quantum limit of the estimand, and that $F_Q/\mathcal{R}_{\rm enc}$ is invariant under adjoining a non-carrier mode.

The test is met at the DCE optimum, where the operating point of \cref{sec:si_energy_budget} carries $n=\num{9.0114}$ photons per mode at the peak. The coherent-state standard quantum limit at that occupancy is $4n=\num{36.05}$~\cite{Caves1981}, and the pure squeezed-vacuum ceiling $F_Q=4n(1+2n)=\num{685.7}$; the DCE value $\num{170903}$ is \num{4741} times the former and \num{249} times the latter. This second ratio serves purely as a scale comparison: the squeezed ceiling applies to phase estimation, whereas the DCE figure estimates the coupling $g$ via parametric modulation, so the two estimands differ. The $1.6\times10^{5}$ ratio quoted in the main text uses as denominator the static two-level model evaluated at $\eta=\num{0.15}$, where that model gives $F_Q=\num{1.069}$, and it must be accompanied by that value and by the statement that the two-level model is not validated at this operating point.

\subsection{Notation}
\label{sec:si_notation_b}

Table~\ref{tab:si_notation_b} fixes the symbols shared with the companion manuscripts~\cite{Deuyekbe2026CompanionGap,Deuyekbe2026CompanionLadder}. The
transmissivity $\eta_{\rm eff}$ of \cref{MAIN-eq:eta_eff} is the only place a damping factor is written
$\eta$: efficiency of a detector, of a conversion stage, or of a displacement readout is always
qualified in words, and $\kappa_{\rm ext}$ is kept distinct from the total linewidth $\kappa$
throughout, since the barrier is a statement about their ratio.

\begin{table}[htbp]
\centering
\caption{Notation used in this work, for symbols that also appear in the companion manuscripts.}
\label{tab:si_notation_b}
{\footnotesize\setlength{\tabcolsep}{3.5pt}
\begin{tabular}{@{}lll@{}}
\toprule
Symbol & Meaning here & Not to be confused with \\
\midrule
$\eta_{\rm eff}$ & transmissivity of the extraction channel & detection efficiency (named in words) \\
$\kappa$, $\kappa_{\rm ext}$ & total and external linewidth & mechanical damping rate \\
$\mathcal{R}_{\rm enc}$ & steady intracavity quanta paid to encode & added noise $n_{\rm add}$ \\
$n_{\rm add}$ & quanta added by the amplifier chain & encoded resource \\
$n_{\rm th}(\omega,T)$ & thermal occupation at the signal frequency & drive-dressing parameter $n_{\rm dr}=g_d^2/\omega_c^2$ \\
$F_Q$ & quantum Fisher information of the probe & record information of a finite sample \\
$\kappa^{*}_{80}$, $\kappa^{*}_{50}$ & linewidth ratio at which retention falls to 80\,\% and 50\,\% & any operating linewidth \\
\bottomrule
\end{tabular}}
\end{table}

\begin{table}[htbp]
\centering
\caption{Parameter and observable glossary for the single-qubit Rabi and collective Dicke protocols ($\hbar = 1$ throughout).}
\label{tab:si_trackB_glossary}
{\small\setlength{\tabcolsep}{4pt}
\begin{tabular}{@{}lll@{}}
\toprule
Symbol & Representative Value / Units & Meaning \\
\midrule
$\omega_c,\,\omega_q,\,\omega_d$ & $\omega_c/2\pi=\qty{6}{\giga\hertz}$ & Cavity, qubit, and modulation frequencies \\
$\eta = g_0/\omega_c$ & \num{0.150} (DCE), \num{1.14}--\num{1.15} (DSC) & Normalized static coupling ratio \\
$g_d/g_0$ & \num{0.11923} (1-qubit), \num{0.15} (Dicke) & Relative DCE modulation amplitude \\
$\kappa/2\pi$ & \qty{246}{\mega\hertz} (modeled), \qty{2.0}{\mega\hertz} ($\kappa^*_{80}$) & Cavity linewidth; \qty{248}{\mega\hertz} is the cited CPW value \\
$F_Q,\,\mathcal{I}(t)$ & Baseline $F_Q^{\rm coh}=\num{1}$, $\mathcal{I}=F_Q/t$ & \parbox[t]{0.40\linewidth}{\raggedright Quantum Fisher information and information rate ($\mathcal{I}$ rate, $R_Q$ retained fraction)} \\
$n_{\rm virt},\,n_{\rm peak}$ & $\lesssim\num{1}$ (static), $\num{9.3}$ (DCE) & Virtual ground-state vs.\ real intracavity photons \\
$N,\,N_{\rm cav}$ & $N\le\num{64}$, $N_{\rm cav}=\num{6}\text{--}\num{24}$ & Qubit count and reduced Dicke Fock truncation \\
\bottomrule
\end{tabular}}
\end{table}

\subsection{Energy budget}
\label{sec:si_energy_budget}

The exact static QFI obeys the reduced budget $F_Q\le 4(2n_{\rm virt}+1)$ with $n_{\rm virt}=\braket{\hat a^\dagger\hat a}_{\Psi_G}$. This bound holds at all 16 points of the regenerated USC sweep, with residuals $4(2n_{\rm virt}+1)-F_Q$ ranging from $\num{+2.50}$ ($\eta=\num{0.7}$) to $\num{+10.39}$ ($\eta=\num{1.3}$). The measured small-$\eta$ asymptote $F_Q\approx 1+3\eta^2$ is consistent with $n_{\rm virt}\approx\eta^2/4$, so the metrological content is carried by the virtual population. Because the single qubit is a two-level system, this budget saturates; the measured curve peaks at $F_Q=\num{6.116}$ at $\eta=\num{1.14}$ and decreases beyond. In the driven DCE protocol at fixed detuning ($\omega_q/\omega_c=\num{1.027}$, $\omega_d/\omega_c=\num{2.077}$, $\eta=\num{0.150}$), sweeping the normalized drive amplitude $g_d/g_0\in[\num{0.05},\num{0.50}]$ exhibits a dual ceiling: the peak QFI rises to $\num{170903}\times$ at $g_d/g_0=\num{0.11923}$ ($n_{\rm peak}=\num{9.01}$) and then collapses under over-driving ($\num{166990}$, $\num{105529}$, $\num{40154}$, and $\num{6131}$ at $g_d/g_0=\num{0.15},\num{0.20},\num{0.30},\num{0.50}$), while the pump-power-normalized ratio $F_Q^{\rm peak}/(g_d/g_0)^2$ falls strictly monotonically across all sampled points from $\num{3.67e7}$ ($g_d/g_0=\num{0.05}$) and $\num{1.20e7}$ ($\num{0.11923}$) down to $\num{2.45e4}$ ($g_d/g_0=\num{0.50}$).

\begin{figure}[htbp]
\centering
\includegraphics[width=\columnwidth]{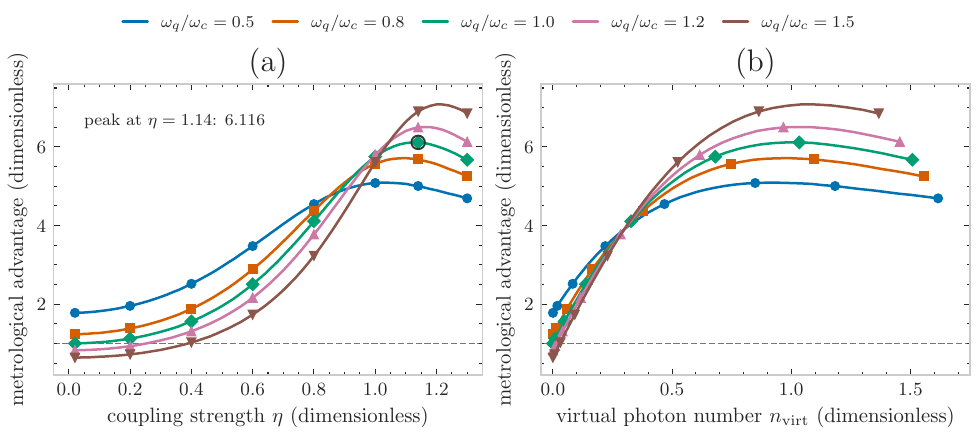}
\caption{Static metrological advantage of the Rabi ground state (QFI relative to the coherent baseline, dimensionless) for $\omega_q/\omega_c\in\{\num{0.5},\num{0.8},\num{1.0},\num{1.2},\num{1.5}\}$. (a) Versus the coupling $\eta$; the resonant curve reaches its sampled maximum \num{6.116} at $\eta=\num{1.14}$. (b) The same data versus the virtual photon number $n_{\rm virt}$, showing that the advantage is organized by the virtual population across detunings.}
\label{fig:si_advantage_landscape}
\end{figure}

\section{Numerical convergence}
\label{sec:si_convergence}

\subsection{Fock truncation}

The Rabi QFI is converged to below $\num{1e-6}$ for $n_{\max}\ge\num{30}$ at $\eta=\num{1.14}$. The Dicke QFI for $N=\num{32}$ is converged to below \qty{0.01}{\percent} for $n_{\max}\ge\num{25}$. The collective spin basis reduces the Hilbert space from $2^N$ to $N+1$ and permits exact diagonalization at $N\le\num{64}$.

\subsection{Time-step and finite-difference convergence of the DCE QFI}

At the quoted single-qubit DCE operating point, the fidelity-based peak QFI varies by less than \qty{0.15}{\percent} across time-step refinements $n_{\rm steps}=\num{100}$--$\num{2000}$ (quoted value at $n_{\rm steps}=\num{500}$). At fixed $n_{\rm steps}=\num{500}$, the peak varies by less than \qty{0.40}{\percent} for $\delta$ between $\num{1e-4}$ and $\num{5e-4}$, and by less than \qty{0.12}{\percent} between $\delta=\num{1e-3}$ (quoted value) and $\delta=\num{1e-4}$. The quoted peak is therefore converged in both parameters.

\section{Single-qubit DCE under cavity loss}
\label{sec:si_loss_adjusted}

\subsection{Loss sweep and retention}

The single-qubit DCE protocol ($\eta=\num{0.150}$, $g_d/g_0=\num{0.11923}$, $\omega_q/\omega_c=\num{1.027}$, $\omega_d/\omega_c=\num{2.077}$) is propagated under the Lindblad master equation with collapse operator $\sqrt{\kappa}\,\hat a$, Fock truncation $N_{\rm cav}=\num{20}$, initial state $\ket{0}\otimes\ket{\downarrow}$, and fidelity-based QFI with $\delta=\num{1e-3}$. \Cref{tab:si_loss_adjusted} reports the peak QFI and its retention versus $\kappa$.

\begin{table}[hbt!]
\centering
\caption{Loss-adjusted single-qubit DCE QFI over the fifteen-point sweep ($N_{\rm cav}=\num{20}$, \num{40} drive periods, $\mathrm{atol}=\num{1e-10}$, $\mathrm{rtol}=\num{1e-8}$). Retention is the peak QFI divided by the pure-state scan maximum \num{170903}.}
\label{tab:si_loss_adjusted}
\begin{tabular}{c c c c}
\toprule
{$\kappa/\omega_c$} & {$\kappa/2\pi$ (MHz)} & {Peak QFI} & {Retention} \\
\midrule
\num{0}      & \num{0}   & \num{170670} & \num{0.9986} \\
\num{1e-5}   & \num{0.06}& \num{169433} & \num{0.9914} \\
\num{5e-5}   & \num{0.3} & \num{165832} & \num{0.9703} \\
\num{1e-4}   & \num{0.6} & \num{161088} & \num{0.9426} \\
\num{2e-4}   & \num{1.2} & \num{151605} & \num{0.8871} \\
\num{3e-4}   & \num{1.8} & \num{142227} & \num{0.8322} \\
\num{5e-4}   & \num{3.0} & \num{124091} & \num{0.7261} \\
\num{7e-4}   & \num{4.2} & \num{107190} & \num{0.6272} \\
\num{1e-3}   & \num{6.0} & \num{84830}  & \num{0.4964} \\
\num{1.5e-3} & \num{9.0} & \num{56125}  & \num{0.3284} \\
\num{3e-3}   & \num{18}  & \num{20133}  & \num{0.1178} \\
\num{5e-3}   & \num{30}  & \num{11664}  & \num{0.0682} \\
\num{1e-2}   & \num{60}  & \num{5946}   & \num{0.0348} \\
\num{2e-2}   & \num{120} & \num{3396}   & \num{0.0199} \\
\num{4.1e-2} & \num{246} & \num{1642}   & \num{0.0096} \\
\end{tabular}
\end{table}

The retention decreases monotonically over the three lowest-loss rows: \qty{99.1}{\percent}, \qty{97.0}{\percent} and \qty{94.3}{\percent} at $\kappa/\omega_c=\num{1e-5}$, \num{5e-5}$ and \num{1e-4}$. No sampled row sits on a plateau. The lossless row itself returns \num{0.9986} of the separately computed pure-state peak used as denominator, which fixes the resolution of the retention axis at about \qty{0.1}{\percent}. At the simulated physical operating point $\kappa/\omega_c=\num{4.1e-2}$ ($\kappa/2\pi\approx\qty{246}{\mega\hertz}$; the cited CPW reference is approximately \qty{248}{\mega\hertz}), retention is $\sim\qty{1}{\percent}$ of the pure-state peak. Retention stays below the bare amplitude-damping envelope $\exp(-\kappa\tau_{\rm int})$ at every sampled loss, by \qty{0.7}{\percent} at the lowest and by a factor \num{24} at the physical point, because the Lindblad dynamics modify the full state evolution and not only the peak amplitude. Retention is therefore a ratio between two computed quantities of the same protocol; the channel-level bounds of the lossy-metrology literature~\cite{Escher2011,Kolodynski2010,Demkowicz2012} refer to a fixed probe transmitted through an attenuation channel and are used here only as the reference against which the driven case is read.

\subsection{Continuous $\kappa$ sweep}
\label{sec:si_kappa_sweep}

A fifteen-point sweep over $\kappa/\omega_c\in[0,\num{4.1e-2}]$, with the seven points $\num{2e-4}$--$\num{1.5e-3}$ added around the knee (\num{500} output samples per point, integrated at $\mathrm{atol}=\num{1e-10}$ and $\mathrm{rtol}=\num{1e-8}$), falls monotonically from $0.991$ at the lowest sampled loss to $0.0096$ at the physical $\kappa$; the full row set is the retention curve of \cref{tab:si_loss_adjusted} and \cref{sec:si_extraction}. The points added on either side of the knee are what confine the fitted exponent: without them the rows sample the crossover from one side only and $\beta$ runs to the edge of its admissible box. Log-linear interpolation gives the crossovers $\kappa^{\ast}_{80}=\num{3.38e-4}$ and $\kappa^{\ast}_{50}=\num{9.89e-4}$, that is \qty{2.0}{\mega\hertz} and \qty{5.9}{\mega\hertz} at a \qty{6}{\giga\hertz} cavity. Both crossovers still fall between sampled points, $80\%$ between \num{3e-4} and \num{5e-4} and $50\%$ between \num{7e-4} and \num{1e-3}, so they are interpolated rather than sampled.

\subsection{Combined cavity, qubit-$T_1$, and temperature row}
\label{sec:si_joint_row}

A single master-equation row combines, at the same operating point, the finite-temperature drive ($T=\qty{20}{\milli\kelvin}$), physical cavity loss ($\kappa/\omega_c=\num{4.1e-2}$), and a conservative qubit-relaxation benchmark ($\gamma_1/\omega_c=\num{1e-4}$), with $N_{\rm cav}=\num{10}$ over $20$ drive periods. The pure-state peak is $\num{30418}$ and the joint peak is $\num{1637}$, an end-to-end retention of \qty{5.4}{\percent}. The joint loss-dominated peak matches the $\kappa$-sweep value at the same $\kappa$ ($\num{1640}$) to within \qty{0.2}{\percent}, confirming cavity decay as the dominant channel. The row is joint in (temperature $\times$ loss $\times$ $T_1$) at the two-level emitter truncation; a multilevel merge remains an open follow-up.

\subsection{Combined cavity and qubit dissipation}
\label{sec:si_qubit_loss}

The qubit coherence computed for this circuit at $\omega_{01}/2\pi=\qty{6.162}{\giga\hertz}$ ($T_1=\qty{25.8}{\micro\second}$, a $Q/\omega_{01}$ bound at dielectric quality factor $Q=\num{1e6}$; $T_\phi=\qty{111}{\micro\second}$ from a $1/f$ flux-noise amplitude $A_\Phi=\num{1e-6}\,\Phi_0$ at $0.01\,\Phi_0$ from the degeneracy point, giving $T_2=\qty{35.3}{\micro\second}$) gives dimensionless rates $\gamma_1/\omega_c=\num{1.03e-6}$ and $\gamma_\phi/\omega_c=\num{2.38e-7}$. Without cavity loss the qubit channels retain \qty{99.75}{\percent} ($T_1$ alone), \qty{99.77}{\percent} ($T_\phi$ alone) and \qty{99.76}{\percent} ($T_1$ and $T_\phi$ together), so the retained fraction tracks the ratio of the interrogation window to $T_2$ rather than the individual rates. At the physical cavity-loss point plus both qubit channels, retention is \qty{0.962}{\percent}, essentially identical to cavity loss alone. Cavity decay is the dominant channel over the tested window.

\section{Collective Dicke+DCE under cavity loss}
\label{sec:si_dicke_loss}

\subsection{Reduced-truncation loss matrix}

The Dicke+DCE operating point ($N=\num{32}$, $\eta=\num{1.15}$, $g_d/g_0=\num{0.15}$, $\omega_q/\omega_c=\num{1.0}$, $\omega_d/\omega_c=\num{2.0}$) is propagated under Lindblad single-photon cavity loss at photon truncations $N_{\rm cav}=\num{6},\num{8},\num{10},\num{12}$ and $\kappa/\omega_c\in\{0,\num{1e-4},\num{1e-3},\num{1e-2},\num{4.1e-2}\}$, with $\texttt{atol}=\texttt{rtol}=\num{1e-10}$ and $n_{\rm steps}=\num{200}$ over ten drive periods. Every stored state satisfies the trace, Hermiticity, and eigenvalue criteria (worst minimum eigenvalue $-\num{1.34e-5}$ against the $-\num{1e-4}$ floor).

\begin{table}[hbt!]
\centering
\caption{Loss-adjusted Dicke+DCE QFI at $N_{\rm cav}=\num{6}$ (Hilbert dim 198). Reference pure-state peak: \num{2962}.}
\label{tab:si_dicke_loss}
\begin{tabular}{c c c c c}
\toprule
{$\kappa/\omega_c$} & {$\kappa/2\pi$ (MHz)} & {Peak QFI} & {Late-window avg} & {Retained} \\
\midrule
\num{0}       & \num{0}     & \num{2962}  & \num{2594} & \qty{100}{\percent} \\
\num{1e-4}   & \num{0.6}   & \num{2927}  & \num{2566} & \qty{99}{\percent} \\
\num{1e-3}   & \num{6}     & \num{2828}  & \num{2490} & \qty{95}{\percent} \\
\num{1e-2}   & \num{60}    & \num{1818}  & \num{1678} & \qty{61}{\percent} \\
\num{4.1e-2} & \num{246}   & \num{474}   & \num{321}  & \qty{16}{\percent} \\
\bottomrule
\end{tabular}
\end{table}

\subsection{Truncation dependence}

Retention is truncation-dependent and not converged. The full trend at the physical $\kappa$ is shown in \cref{tab:si_dicke_loss_convergence}. The pure-state values quoted in the main text (late-window average $\num{167271}$ and boundary peak $\num{251557}$ at period \num{40}) come from the production propagation at $N_{\rm cav}=\num{28}$; at $N_{\rm cav}=\num{6}$ the same protocol gives a boundary peak of $\num{33109}$ at period \num{40} and $\num{2962}$ at period \num{10}, and the latter is the zero-loss reference of the retentions in \cref{tab:si_dicke_loss}.

\begin{table}[hbt!]
\centering
\caption{Truncation dependence of Dicke+DCE QFI retention at the physical cavity linewidth. Retention is relative to each truncation's own zero-loss peak.}
\label{tab:si_dicke_loss_convergence}
\begin{tabular}{c c c c c}
\toprule
{$N_{\rm cav}$} & {Hilbert dim} & {Peak QFI ($\kappa=0$)} & {Peak QFI (physical $\kappa$)} & {Retention} \\
\midrule
\num{6}  & \num{198} & \num{2962}  & \num{474}  & \qty{16}{\percent} \\
\num{10} & \num{330} & \num{2363}  & \num{290}  & \qty{12}{\percent} \\
\num{12} & \num{396} & \num{4833}  & \num{320}  & \qty{7}{\percent}  \\
\num{16} & \num{528} & \num{5786}  & \num{255}  & \qty{4.4}{\percent} \\
\num{20} & \num{660} & \num{9851}  & \num{265}  & \qty{2.7}{\percent} \\
\num{24} & \num{792} & \num{10584} & \num{273}  & \qty{2.6}{\percent} \\
\bottomrule
\end{tabular}
\end{table}

The pure-state peak is non-monotonic across the six bases (\cref{tab:si_dicke_loss_convergence}), reflecting Fock-space confinement at the smallest truncation. Retention decreases monotonically. Of the six truncation comparisons against the larger truncation ($N_{\rm cav}=\num{16}\to\num{24}$ and $\num{20}\to\num{24}$ at each of the three loss values), two meet the \qty{5}{\percent} tolerance ($\num{16}\to\num{24}$ at $\kappa/\omega_c=\num{1e-2}$, $\num{20}\to\num{24}$ at the physical $\kappa$); the rest exceed it. The physical-point retention is therefore a truncation-dependent range of roughly \qty{16}{\percent} ($N_{\rm cav}=\num{6}$) down to \qty{2.6}{\percent} ($N_{\rm cav}=\num{24}$), not a converged number.

Two conventions keep the retentions comparable across rows. First, retention is always a ratio taken at matched interrogation window: the lossy run and its lossless reference are propagated for the same number of drive periods, and the reference is the same family at the same truncation, so no row divides a ten-period peak by a forty-period one. Second, the absolute peak is never quoted as converged, since \cref{sec:si_dicke_convergence} shows it growing monotonically with the window; the retention therefore inherits the window, and the numbers reported are the ones at the stated windows. Within one family the accumulation of loss also moves the interrogated optimum: the collective retention peak sits at \num{9.65} drive periods at zero loss and at \num{6.13} periods at $\kappa/\omega_c=\num{4.1e-2}$, so loss shortens the interrogation that would extract the information as well as shrinking what is extractable.

\subsection{Boundary-peak convergence}
\label{sec:si_dicke_convergence}

Extending the pure-state Dicke+DCE time trace at $N_{\rm cav}=\num{6}$ to periods $\num{20}$--$\num{60}$ shows the peak QFI continues to grow monotonically and occurs at the last stored time point for every window (\cref{tab:si_dicke_convergence}). Both the production boundary value $\num{251557}\times$ ($N_{\rm cav}=\num{28}$, $T_{\max}=\num{40}$) and the $N_{\rm cav}=\num{6}$ value $\num{33109}$ at $T_{\max}=\num{40}$ therefore reflect the finite simulation cutoff rather than converged local maxima.

\begin{table}[hbt!]
\centering
\caption{Dicke+DCE boundary-peak convergence at $N_{\rm cav}=\num{6}$. The peak occurs at the last stored time point for every window.}
\label{tab:si_dicke_convergence}
\begin{tabular}{c c c}
\toprule
{$T_{\rm max}$ (periods)} & {Peak QFI} & {Late-window avg} \\
\midrule
\num{20} & \num{9932}  & \num{8219}  \\
\num{30} & \num{19093} & \num{16060} \\
\num{40} & \num{33109} & \num{27535} \\
\num{50} & \num{48328} & \num{39760} \\
\num{60} & \num{69475} & \num{57117} \\
\bottomrule
\end{tabular}
\end{table}

\subsection{Photon-sector decomposition}

Replaying the validated Dicke+DCE rows and projecting onto the $n\le n_{\rm cut}$ sector shows the low-photon blocks carry the parameter-relevant coherence. At $N_{\rm cav}=\num{6}$ ($\langle n\rangle=\num{2.9}$--$\num{3.5}$) the $n\le\num{2}$ sector peak exceeds the full-space peak ($\num{2.65}\times$ at $\kappa=0$, $\num{1.28}\times$ at the physical $\kappa$) while retaining only \qty{33}{\percent}--\qty{37}{\percent} trace weight; the overshoot decreases monotonically to $1.000$ as $n_{\rm cut}\to N_{\rm cav}$. At $N_{\rm cav}=\num{12}$ the sector curve converges monotonically with the cutoff (\cref{tab:si_photon_sector}), and at the physical $\kappa$ the $n_{\rm cut}=\num{10}$ projection (\qty{84}{\percent} weight) is within \qty{0.1}{\percent} of the full peak.

\begin{table}[hbt!]
\centering
\caption{Fraction of the full-space peak QFI retained by the $n\le n_{\rm cut}$ photon sector at $N_{\rm cav}=\num{12}$; trace weight in parentheses.}
\label{tab:si_photon_sector}
\begin{tabular}{lrrrrr}
\toprule
$n_{\rm cut}$ & \num{4} & \num{6} & \num{8} & \num{10} & \num{12} \\
\midrule
$\kappa/\omega_c=\num{1e-2}$ & $0.791\,(0.34)$ & $0.860\,(0.51)$ & $0.897\,(0.67)$ & $0.946\,(0.83)$ & $1.000\,(1.00)$ \\
$\kappa/\omega_c=\num{4.1e-2}$ & $0.753\,(0.38)$ & $0.925\,(0.54)$ & $0.988\,(0.69)$ & $1.001\,(0.84)$ & $1.000\,(1.00)$ \\
\bottomrule
\end{tabular}
\end{table}

Sector QFI is not additive across sectors: under tight truncation ($N_{\rm cav}=\num{6}$), reflection off the upper Fock boundary artificially compresses curvature into the low-photon subspace and produces the overshoot reported above at $n_{\rm cut}=\num{2}$, explaining why reduced bases overstate dissipative retention relative to converged truncations. The curves serve as a truncation-sensitivity diagnostic rather than a decomposition of the full QFI into independent contributions.

\section{Cavity-channel HEOM cross-check}
\label{sec:si_cavity_heom}

The Markovian cavity-loss treatment underlying the retention values in \cref{sec:si_loss_adjusted,sec:si_dicke_loss} is cross-checked against a cavity-mode HEOM calculation: a Fock-space mode ($d=\num{12}$) coupled through the position quadrature $\hat x=\hat a+\hat a^\dagger$ to a structured Drude--Lorentz bath and benchmarked against $\sqrt{\kappa}\hat a$ Lindblad. The grid spans $\kappa/\omega_c\in\{\num{1e-4},\num{4.1e-2}\}$, Drude cutoffs $\gamma/\omega_c\in\{\num{5},\num{10},\num{20},\num{40}\}$, hierarchy depths $L=\num{2},\num{3},\num{4}$, temperatures $T=\num{0.01},\num{0.05}$, and windows $t_{\rm final}=\num{1},\num{2}$ (\num{96} rows).

At $\kappa/\omega_c=\num{1e-4}$ all \num{18} physical rows give mean fidelity errors of $\num{9.2e-5}$--$\num{3.7e-4}$; the hierarchy shows no depth dependence, confirming that Markovian and non-Markovian dynamics are indistinguishable at this linewidth. At $\kappa/\omega_c=\num{4.1e-2}$ the \num{18} physical rows give mean state infidelities of \qty{4.8}{\percent}--\qty{24}{\percent}; the discrepancy grows with the Drude cutoff and with the observation window, from \qty{4.8}{\percent} at $\gamma/\omega_c=\num{10}$ over the short window to \qty{24}{\percent} at $\gamma/\omega_c=\num{40}$ over the long one, and the hierarchy converges at depth $L=\num{3}$. The \num{60} remaining rows fail the positivity criterion (minimum eigenvalue down to $-\num{1.4}$), documenting a genuine numerical limitation of the HEOM hierarchy for strong cavity-bath coupling.

At the physical linewidth the Lindblad retentions are therefore first-order Markovian estimates, and the \qtyrange{4.8}{24}{\percent} infidelity range is a model-comparison envelope for cavity-bath memory, not a statistical error bar. The rows do not resolve how bath memory deforms the knee $x_c$ of the extraction law; that question requires retention-level HEOM runs.

\section{Multilevel-emitter leakage}
\label{sec:si_multilevel}

The two-level approximation used in the main-text QFI computations is tested against a common oscillator-based emitter interface parameterized by declared transition-frequency ratios for three effective spectra: transmon-like, flux-qubit-like, and fluxonium-like. Each retains $d_e=\num{2},\dots,\num{5}$ emitter levels. Static rows use $\eta=\num{0.15}$ and $\num{1.0}$; driven rows use drive coefficients $g_d=\num{0.11923}$ (reference) and $\num{0.5}$ (strong) at $\kappa/\omega_c=0$ and $\num{4.1e-2}$ over \num{40} drive periods. The adequacy criterion is declared before scanning: maximum leakage to levels $\ket{2}$ and above below \qty{1}{\percent} of the population over the window.

No driven $d_e\ge\num{3}$ row meets the criterion at the reference drive. Window-maximum leakage ranges from \qty{2.99}{\percent} (fluxonium-like, $d_e=\num{3}$) to \qty{81.5}{\percent} (fluxonium-like, $d_e=\num{5}$, strong drive). Paired two-level versus multilevel comparisons show the multilevel peak QFI relative change from the two-level baseline ranging from $-0.873$ to $+7.36$, with the largest positive deviations at strong drive where leakage is largest.

A staged drive-amplitude sweep over $g_d\in\{0,\num{0.02},\num{0.04},\num{0.06},\num{0.08},\num{0.10},\num{0.11923}\}$ at $d_e=\num{3}$ (\num{42} rows, $\kappa/\omega_c=0$ and $\num{4.1e-2}$) identifies platform-specific safe operating windows at physical loss (\cref{tab:si_multilevel_sweep}). The QFI enhancement at those leakage-acceptable points is orders of magnitude below the reference value.

\begin{table}[hbt!]
\centering
\caption{Maximum drive coefficient $g_d$ satisfying the \qty{1}{\percent} leakage ceiling at physical loss ($d_e=\num{3}$, \num{40} drive periods) and the corresponding peak QFI for the coupling estimand.}
\label{tab:si_multilevel_sweep}
\begin{tabular}{l c c}
\toprule
{Platform} & {Max $g_d$} & {Peak QFI} \\
\midrule
Fluxonium-like & $\num{0.06}$ & $\num{85.5}$ \\
Flux-qubit-like & $\num{0.04}$ & $\num{77.5}$ \\
Transmon-like & $\num{0.02}$ & $\num{54.2}$ \\
\bottomrule
\end{tabular}
\end{table}

The trade-off between leakage and QFI enhancement is monotonic. The criterion is applied to the whole trajectory rather than to the instant of the QFI peak, since a transient excursion above the threshold breaks the two-level truncation as surely as a sustained one: read that way, \num{18} of the \num{42} screened rows at $d_e=\num{3}$ pass, and at physical loss the largest admissible amplitude is the one tabulated above. At the amplitude that produces the headline gain, $g_d=\num{0.11923}$, all six model--loss combinations fail, at \qty{3.0}{\percent} to \qty{11.9}{\percent}. The \qty{1}{\percent} leakage ceiling is a model choice, and the models are declared effective spectra, not device simulations.

\section{Finite-size scaling of the Dicke model}
\label{sec:si_finite_size}

The exact static Dicke ground-state QFI is evaluated at two cavity truncations per size bracket for $N=\num{4},\dots,\num{64}$ at $\omega_q/\omega_c=\num{1.0}$ and $\num{1.005}$ (\num{52} rows). All static rows are truncation-converged (relative change below \qty{5}{\percent}; typical values $10^{-11}$--$10^{-14}$). The static QFI at $N=\num{32}$, $\omega_q/\omega_c=\num{1.0}$ is \num{228.6}, reproducing the reference value.

Descriptive power-law fits over $N\in[\num{6},\num{18}]$ give $\alpha=\num{2.197}\pm\num{0.039}$ ($\omega_q/\omega_c=\num{1.0}$, $R^2=\num{0.9984}$) and $\alpha=\num{2.216}\pm\num{0.034}$ ($\omega_q/\omega_c=\num{1.005}$, $R^2=\num{0.9988}$), consistent with the historical $\alpha=\num{2.21}$. Dropping the largest in-range point changes $\alpha$ by only \num{0.004}. Beyond $N\approx\num{21}$ the local growth exponent falls below unity (\num{0.67} at $N=\num{24}$, \num{0.34} at $N=\num{27}$) before recovering to \num{0.85} at $N=\num{64}$. Single-plateau saturated forms do not describe the data over $N\in[\num{4},\num{64}]$, so no saturation plateau is reached up to $N=\num{64}$ within this grid. The fits are recorded with uncertainty intervals and boundary sensitivity; they are explicitly descriptive.

Driven Dicke+DCE rows at $N_{\rm cav}=\num{6}$ show the lossless peak growing with $N$ at a fixed pump proxy $g_d^2=\num{0.0298}$, while cavity loss reduces the peak and moves it into the window (\cref{tab:si_finite_size}).

\begin{table}[hbt!]
\centering
\caption{Driven Dicke+DCE resource-accounting at $g_d^2=\num{0.0298}$ ($N_{\rm cav}=\num{6}$). Emitted proxy is $\kappa\int n(t)\,dt$, not a measured output.}
\label{tab:si_finite_size}
\begin{tabular}{c c c c c c}
\toprule
{$N$} & {$\kappa/\omega_c$} & {Peak $F_Q$} & {Peak period} & {Max $\bar n$} & {Emitted proxy} \\
\midrule
\num{4}  & \num{0}      & \num{785.3} & \num{10.0} & \num{3.25} & \num{0}     \\
\num{4}  & \num{1e-2}   & \num{469.3} & \num{10.0} & \num{3.17} & \num{0.663} \\
\num{4}  & \num{4.1e-2} & \num{96.0}  & \num{6.8}  & \num{3.06} & \num{2.394} \\
\num{12} & \num{0}      & \num{1851.6}& \num{10.0} & \num{3.19} & \num{0}     \\
\num{12} & \num{1e-2}   & \num{1072.6}& \num{10.0} & \num{2.90} & \num{0.696} \\
\num{12} & \num{4.1e-2} & \num{188.6} & \num{7.7}  & \num{2.70} & \num{2.830} \\
\num{24} & \num{0}      & \num{2768.1}& \num{10.0} & \num{2.94} & \num{0}     \\
\bottomrule
\end{tabular}
\end{table}

The screen supports a descriptive superlinear finite-size trend over the declared interval, with uncertainty and boundary sensitivity reported; establishing an empirical small-$N$ scaling trend rather than an asymptotic critical exponent, while the emitted-photon column records the integrated cavity loss $\kappa\int n(t)\,dt$ rather than a jointly optimized resource ratio.

Companion eta-sweep curves at $N=\num{4},\num{8},\num{16},\num{32}$ (produced by the same physics core and anchored at $\eta=\num{1.15}$ against the finite-size grid to below $\num{3e-7}$ relative deviation) show the sampled static-QFI maximum shifting with $N$ ($\approx\num{1.40}$ at $N=\num{4}$, $\approx\num{1.15}$ at $N=\num{16}$, $\approx\num{1.10}$ at $N=\num{32}$). The tabulated $\eta=\num{1.15}$ operating point is therefore not the sampled optimum for every $N$.

\section{Additional model-specific screens}
\label{sec:si_additional}

The following screens bound the operational significance of the main-text QFI values but are not the central results. They are grouped here to keep the main text focused.

\subsection{Thermal-competition screen}
\label{sec:si_thermal}

A finite-temperature Lindblad screen over $T\in\qtyrange{10}{200}{\milli\kelvin}$ and $\omega_c/2\pi\in[\num{5},\num{8}]~\unit{\giga\hertz}$ uses explicit emission and absorption channels $L_-=\sqrt{\kappa(n_{\rm th}+1)}\,\hat a$, $L_+=\sqrt{\kappa n_{\rm th}}\,\hat a^\dagger$ with $n_{\rm th}$ evaluated at the declared cavity frequency, plus separate qubit-thermalization rows. The retention criterion is the first sampled temperature at which peak QFI relative to the same-configuration zero-temperature reference falls below \qty{80}{\percent}.

For the reference driven row ($\eta=\num{0.15}$, vacuum initial state, $\kappa/\omega_c=\num{4.1e-2}$, $\omega_c/2\pi=\qty{6}{\giga\hertz}$), the zero-temperature peak QFI is \num{1642.1}, retention is \num{0.985} at \qty{50}{\milli\kelvin} and \num{0.775} at \qty{100}{\milli\kelvin}, so the sampled crossover is at \qty{100}{\milli\kelvin}. The cavity Bose--Einstein occupation is $\num{5.6e-7}$ at \qty{20}{\milli\kelvin} and \num{5.95e-2} at \qty{100}{\milli\kelvin}. The screen is fixing a sampled crossover temperature within the tested cavity frequency and bath model.

\subsection{Homodyne output-field CFI screen}
\label{sec:si_homodyne}

An idealized joint-system QFI does not by itself specify a laboratory readout. A bounded output-field screen uses the input--output convention $a_{\rm out}(t)=a_{\rm in}(t)+\sqrt{\kappa_{\rm ext}}a(t)$ with vacuum input and $\kappa=\kappa_{\rm ext}+\kappa_{\rm int}$; only the external fraction is treated as monitored output. The measured temporal mode is flat over \num{6} or \num{10} drive periods, with a Gaussian quadrature distribution. Among rows with a valid same-state comparison, $F_C/F_Q$ ranges from $0$ to \num{0.603}; the largest static ratio is \num{0.0455}, falling below the \qty{92.4}{\percent} (static) and \qty{88.1}{\percent} (driven) ratios reported for idealized closed-encoding benchmarks; optimal matched-filter homodyne schemes are left for future study.

\subsection{Multiparameter QFI and representation consistency}
\label{sec:si_multiparameter}

Directional sensitivity is evaluated for the parameter vector $(\eta,\omega_q/\omega_c,g_d/g_0,\phi)$ in the driven protocol and $(\eta,\omega_q/\omega_c)$ for static ground states, using phase-aligned central finite differences at steps $\num{1e-2}$--$\num{1e-4}$. The nuisance-projected information reported below is the Schur complement of the multiparameter Fisher matrix with respect to the parameters retained, so a large diagonal entry can coexist with a small projected one when the optimal measurements are incompatible~\cite{Ragy2016,Szczykulska2016}. All \num{18} rows are finite, positive semidefinite within tolerance, and converged. In the static USC ground state, virtual dressing renders $\eta$ and $\omega_q/\omega_c$ nearly collinear: normalized correlation is $\num{-0.99755}$ at $\eta=\num{0.10}$ (condition number $\num{8.63e4}$) and $\num{-0.99460}$ at $\eta=\num{0.15}$ ($\num{1.81e4}$), collapsing nuisance-projected $F_{\eta,{\rm eff}}$ to $\num{0.0049}$ of $\num{1.0032}$ (\qty{0.49}{\percent}) and $\num{0.0112}$ of $\num{1.0411}$ (\qty{1.08}{\percent}), respectively. Entering DSC ($\eta=\num{1.0}$) relaxes the correlation to $\num{-0.89461}$ and drops the condition number to $\num{175.3}$, lifting retained information to \qty{20.0}{\percent} ($F_{\eta,{\rm eff}}=\num{1.1512}$ of $\num{5.7657}$). Floquet DCE driving at $\eta=\num{0.15}$ similarly breaks the static degeneracy ($\rho_{\eta,\omega_q}\in[\num{0.431},\num{0.823}]$, condition numbers $\num{794}\text{--}\num{1165}$, retained fraction \qty{23}{\percent}--\qty{50}{\percent}); at the reference \num{10}-period row and $N_{\rm cav}=\num{20}$, the coupling diagonal is approximately $\num{1.15e3}$ and the nuisance-projected effective coupling information is approximately $\num{4.08e2}$.

A controlled similarity transformation $U=\exp[-i\num{0.37}a^\dagger a]$ at $N_{\rm cav}=\num{30}$ transforms the Hamiltonian, initial state, control, and observables together; unitarity error is \num{1.6e-15}, transformed-state infidelity below \num{4e-16}, and the relative QFI-metric difference is \num{4.8e-9}. confirming finite-dimensional similarity covariance for a shared control parameterization.

\subsection{Interrogation-window crossover}
\label{sec:si_interrogation}

At the fixed single-qubit DCE operating point, the fidelity-based QFI and the information-rate objective $\mathcal{I}(t)=F_Q(t)/t$ are evaluated over \num{40} drive periods for $\kappa/\omega_c\in\{0,\num{1e-4},\num{1e-3},\num{5e-3},\num{1e-2},\num{2e-2},\num{4.1e-2}\}$. At $\kappa/\omega_c\le\num{1e-3}$ both metrics remain boundary-limited; at larger loss the $\mathcal{I}$ maximum moves earlier, from \num{18.08} to \num{6.96} drive periods between $\kappa/\omega_c=\num{0.005}$ and $\num{4.1e-2}$, while the peak $F_Q$ falls from \num{10449} to \num{1242} and the photon occupation from \num{5.08} to \num{0.76} (\cref{tab:si_interrogation}). Across that entire interior-optimum range, however, the probe von Neumann entropy at $\mathcal{I}_{\max}$ remains locked within $S\in[\num{1.35},\num{1.60}]$ bits, while the peak-QFI time stabilizes around $S\in[\num{1.80},\num{2.05}]$ bits, showing that stronger dissipation reaches the same mixedness ceiling earlier rather than tolerating higher entropy. The stored entropy derivative serves as an empirical finite-difference mixedness diagnostic within the driven Rabi model rather than a universal thermodynamic bound.

\begin{table}[hbt!]
\centering
\caption{Information-rate maxima for the single-qubit DCE protocol.}
\label{tab:si_interrogation}
\begin{tabular}{c c c c c}
\toprule
{$\kappa/\omega_c$} & {$\kappa/2\pi$ (MHz)} & {$t/T_d$ at $\mathcal{I}$ max} & {$F_Q$ at max} & {Window status} \\
\midrule
\num{0}      & \num{0}   & \num{40.00} & \num{170670} & boundary \\
\num{1e-4}   & \num{0.6} & \num{40.00} & \num{161087} & boundary \\
\num{1e-3}   & \num{6}   & \num{40.00} & \num{84830}  & boundary \\
\num{5e-3}   & \num{30}  & \num{18.08} & \num{10449}  & interior \\
\num{1e-2}   & \num{60}  & \num{14.00} & \num{5510}   & interior \\
\num{2e-2}   & \num{120} & \num{10.96} & \num{2961}   & interior \\
\num{4.1e-2} & \num{246} & \num{6.96}  & \num{1242}   & interior \\
\bottomrule
\end{tabular}
\end{table}

\subsection{Matched-photon and adaptive-protocol references}
\label{sec:si_matched}

At matched intracavity photon budget $n=\num{9.01}$ (the DCE peak), the pure-state DCE QFI is $\num{4741}\times$ the coherent-state shot-noise reference $F_Q=4n$ and $\num{249}\times$ the squeezed-state reference $F_Q=4n(1+2n)$. Photon-number matching leaves pump power, bandwidth, preparation, loss, and readout unmatched, so these ratios calibrate the pure-state ceiling rather than device-level advantage. Under cavity loss the DCE QFI collapses as quantified in \cref{sec:si_loss_adjusted}.

A formal comparison with adaptive protocols is retained only as a reference at the DCE peak photon budget and the DCE time-average budget. The adaptive entries are uncertainty-bound or formal-reference expressions, whereas the DCE entries are QFI values, so no cross-estimand enhancement ratios are reported. A physically relevant comparison would require a common estimation task, prior, observation time, loss model, drive budget, and readout; that comparison is outside the present study.

\subsection{Finite-window Floquet-stability screen}

A deterministic screen at the driven single-qubit Rabi model ($\omega_q/\omega_c=\num{1.027}$, $g(t)=g_0[1+(g_d/g_0)\cos(\omega_d t+\phi)]$, cavity loss $\kappa$) evaluates coupling-estimand QFI from paired trajectories at $\eta$ and $\eta+\delta\eta$, with $\delta\eta=\num{1e-3}$. Tier 1 uses $N_{\rm cav}=\num{12},\num{20}$ and \num{40} periods over \num{108} parameter rows; Tier 2 validates five representative rows at \num{80} periods and $N_{\rm cav}=\num{20},\num{30}$. All trajectories satisfy the trace, Hermiticity, and eigenvalue criteria.

Tier 1 classifies \num{79} rows as showing boundary growth and \num{29} as bounded within the screen; no row exceeds the \qty{10}{\percent} truncation-difference criterion. Tier 2 resonant/low-loss rows continue to grow at the boundary, whereas two selected lossy rows remain bounded and agree between the two truncations to better than $10^{-6}$ in peak QFI. The persistence of boundary growth in low-loss resonant rows rejects the claim that two-level saturation is a general passive Floquet stabilizer. The screen is finite-window; no asymptotic stability or phase stability is established.

\subsection{HEOM cross-checks (qubit bath)}
\label{sec:si_heom}

Two HEOM comparisons bound the Lindblad working model for the qubit-bath channel.

\textit{Local short-time check.} A Drude--Lorentz bath ($\omega_q=\num{1.0}$, $\gamma_{\rm cut}=\num{20}$, $\lambda=\num{0.02}$, $\Gamma=\num{0.012535}$, $T=\num{0.001}$, hierarchy depth \num{3}, \num{50} time points to $t_{\rm final}=\num{1.0}$) gives a mean Uhlmann-fidelity error of \qty{1.18}{\percent} (maximum \qty{2.43}{\percent}), all sampled points below \qty{5}{\percent}.

\textit{Extended check.} A hierarchy-depth, temperature, coupling, cutoff, and window sweep (\num{18} rows, \num{15} physical, \num{3} inconclusive) confirms hierarchy convergence at $L=\num{2}$ (relative change \num{4.3e-6} from $L=\num{2}$ to \num{5}). Baseline agreement is \num{5.3e-4} mean, \num{6.6e-4} maximum. Two driven DCE-ratio rows give QFI relative differences of \qty{1.05}{\percent} and \qty{1.09}{\percent} at $\eta_d=\num{0.15}$ and \num{0.10}. The $T=\num{0.35}$ row shows a larger mean error (\qty{7.7}{\percent}), retained as a temperature-scope caveat.

Neither HEOM check validates the driven high-QFI traces; both support the working Lindblad model for their declared scope only.

\section{Experimental feasibility}
\label{sec:si_feasibility_B}

\Cref{tab:si_experimental_feasibility} compares the model target with the demonstrated state of the art at the key parameters.

\begin{table}[hbt!]
\centering
\caption{Experimental feasibility assessment.}
\label{tab:si_experimental_feasibility}
\footnotesize
\setlength{\tabcolsep}{4pt}
\begin{tabular}{l c c c}
\toprule
\textbf{Requirement} & \textbf{Model target} & \textbf{CPW (current USC)} & \textbf{3D cavity (candidate)} \\
\midrule
    \parbox[t]{3.2cm}{\raggedright Number of qubits $N$} & \parbox[t]{3.6cm}{\raggedright \num{32}} & \parbox[t]{4.0cm}{\raggedright \num{4} \cite{Yoshihara2017}} & \parbox[t]{4.0cm}{\raggedright \num{1}~\cite{Paik2011}} \\
    \parbox[t]{3.2cm}{\raggedright Coupling $\eta$} & \parbox[t]{3.6cm}{\raggedright \num{1.15}} & \parbox[t]{4.0cm}{\raggedright \num{1.3} \cite{Niemczyk2010}} & \parbox[t]{4.0cm}{\raggedright dispersive~\cite{Paik2011}} \\
    \parbox[t]{3.2cm}{\raggedright Coupling $g/2\pi$} & \parbox[t]{3.6cm}{\raggedright \qty{6.9}{\giga\hertz}} & \parbox[t]{4.0cm}{\raggedright \qty{6}{\giga\hertz} \cite{Niemczyk2010}} & \parbox[t]{4.0cm}{\raggedright ---} \\
    \parbox[t]{3.2cm}{\raggedright DCE drive $g_d/2\pi$} & \parbox[t]{3.6cm}{\raggedright \qty{1.0}{\giga\hertz}} & \parbox[t]{4.0cm}{\raggedright Tunable coupler~\cite{Niemczyk2010}; not at $g_d/g_0=\num{0.12}$} & \parbox[t]{4.0cm}{\raggedright Requires dedicated pump port} \\
    \parbox[t]{3.2cm}{\raggedright Drive frequency $\omega_d/2\pi$} & \parbox[t]{3.6cm}{\raggedright \qty{12}{\giga\hertz}} & \parbox[t]{4.0cm}{\raggedright Standard} & \parbox[t]{4.0cm}{\raggedright Standard} \\
    \parbox[t]{3.2cm}{\raggedright Temperature} & \parbox[t]{3.6cm}{\raggedright $<\qty{200}{\milli\kelvin}$} & \parbox[t]{4.0cm}{\raggedright \qty{10}{\milli\kelvin}} & \parbox[t]{4.0cm}{\raggedright \qty{10}{\milli\kelvin}} \\
    \parbox[t]{3.2cm}{\raggedright Cavity $\kappa/2\pi$} & \parbox[t]{3.6cm}{\raggedright $<\qty{1}{\mega\hertz}$} & \parbox[t]{4.0cm}{\raggedright \qty{248}{\mega\hertz}} & \parbox[t]{4.0cm}{\raggedright \qty{0.1}{\mega\hertz}~\cite{Paik2011}} \\
    \parbox[t]{3.2cm}{\raggedright Coherence $T_1$} & \parbox[t]{3.6cm}{\raggedright $\gg\qty{83}{\nano\second}$} & \parbox[t]{4.0cm}{\raggedright \qtyrange{10}{100}{\micro\second}} & \parbox[t]{4.0cm}{\raggedright \qtyrange{10}{100}{\micro\second}} \\
    \parbox[t]{3.2cm}{\raggedright Floquet bound-state rescue} & \parbox[t]{3.6cm}{\raggedright demonstrable at $\kappa/\omega\sim\num{4e-2}$} & \parbox[t]{4.0cm}{\raggedright emitter-local only~\cite{Bai2023}} & \parbox[t]{4.0cm}{\raggedright unshown for cavity leakage} \\
\bottomrule
\end{tabular}
\end{table}

No demonstrated platform simultaneously satisfies (i) the $\eta=\num{1.15}$ coupling needed for the superradiant collective enhancement, (ii) the sampled single-qubit reference point $\kappa/2\pi=\qty{0.6}{\mega\hertz}$ at which the loss sweep retains \qty{94.3}{\percent}, and (iii) the uniform collective coupling of $N=\num{32}$ qubits. The cited CPW example provides strong coupling and a partial collective architecture at $\kappa/2\pi\approx\qty{248}{\mega\hertz}$; the cited 3D-cavity example provides the low-linewidth regime at the single-qubit level.

\begin{figure}[hbt!]
\centering
\includegraphics[width=0.7\columnwidth]{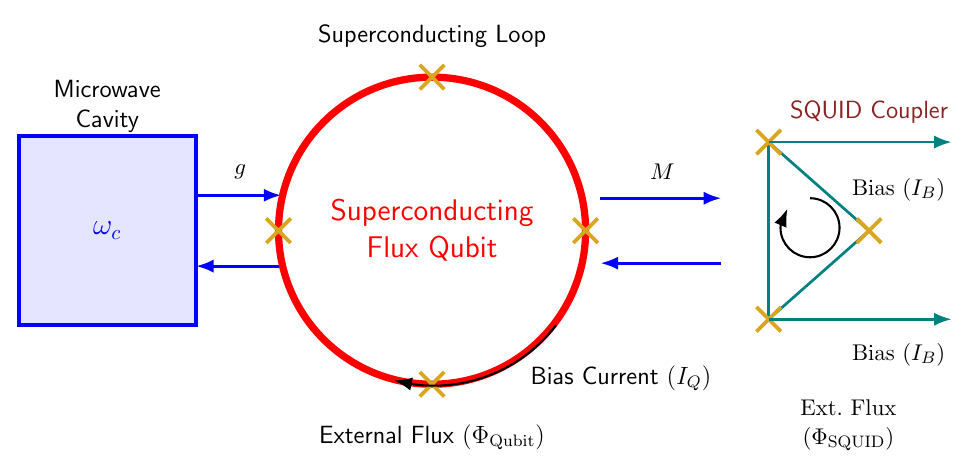}
\caption{Flux qubit platform. A superconducting loop with Josephson junctions forms a flux qubit (red) coupled to a microwave cavity (blue); external flux $\Phi_{\text{ext}}$ tunes the qubit frequency.}
\label{fig:platform_flux_qubit}
\end{figure}

\begin{figure}[hbt!]
\centering
\includegraphics[width=0.7\columnwidth]{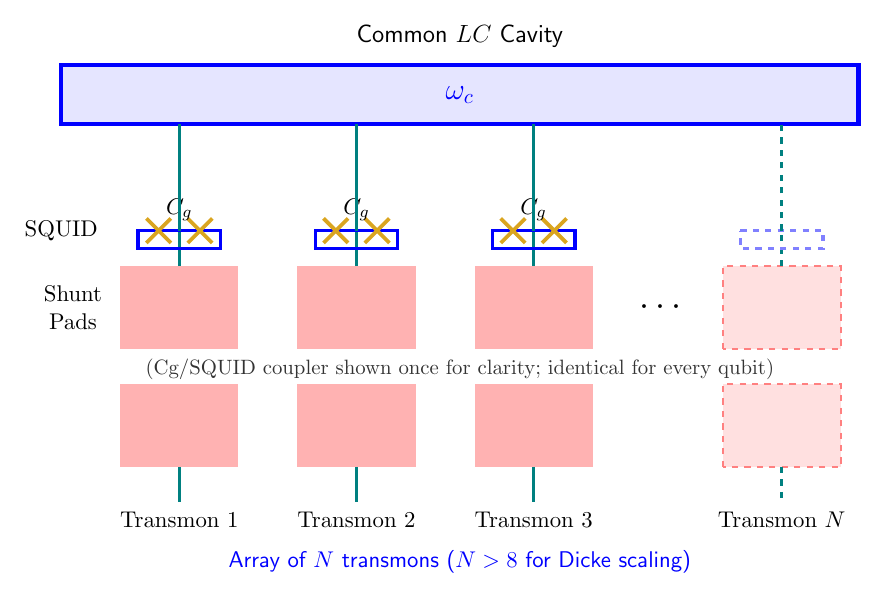}
\caption{Transmon platform. An array of transmon qubits (red) capacitively coupled to a common $LC$ cavity (blue) via tunable couplers $C_g$.}
\label{fig:platform_transmon}
\end{figure}

\begin{figure}[hbt!]
\centering
\includegraphics[width=0.7\columnwidth]{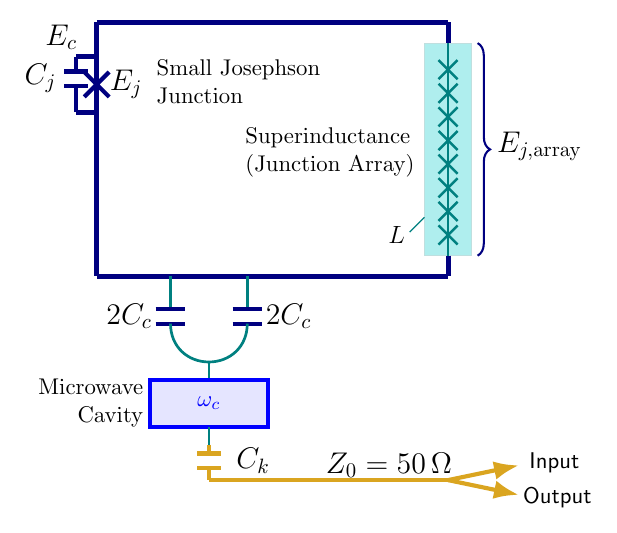}
\caption{Fluxonium platform. High anharmonicity and long coherence enable operation at $\eta\sim\num{1.3}$ with $T_2>\qty{10}{\micro\second}$.}
\label{fig:platform_fluxonium}
\end{figure}

\textit{Platform candidates.} Flux qubits are the original USC platform with tunable coupling, best for $N=\numrange{1}{4}$~\cite{Niemczyk2010}. Transmons are less anharmonic but more coherent, reaching USC with flux-tunable couplers, best for $N>\num{8}$ due to fabrication uniformity~\cite{Yoshihara2017}. Fluxonium offers high anharmonicity and long coherence, promising for USC with $N=\numrange{1}{2}$~\cite{Kockum2019}. The three platform schematics are shown in \cref{fig:platform_flux_qubit,fig:platform_transmon,fig:platform_fluxonium}.

\textit{Staged program.} The experimental roadmap has three stages. Stage~1 ($N=\num{1}$, single qubit, flux-based CPW or 3D cavity) benchmarks the cavity-loss-augmented QFI against the single-qubit retention curve of \cref{sec:si_loss_adjusted}; a 3D cavity at $\omega_c/2\pi=\qty{6}{\mega\hertz}$ is a candidate platform. Stage~2 ($N=\numrange{4}{8}$) tests the collective scaling exponent in the loss-controlled regime at the quoted $\eta=\num{1.15}$, without engaging the DCE modulation. Stage~3 ($N=\numrange{16}{32}$) simultaneously demonstrates collective dressing, DCE modulation, and the loss barrier; a 3D-cavity-based array is a candidate platform for the required low linewidth, but whether it can also provide the required collective coupling uniformity remains open. None of the three stages has been demonstrated end-to-end.

\subsection{Parameter-dependence scan}

\begin{figure}[hbt!]
\centering
\includegraphics[width=.97\columnwidth]{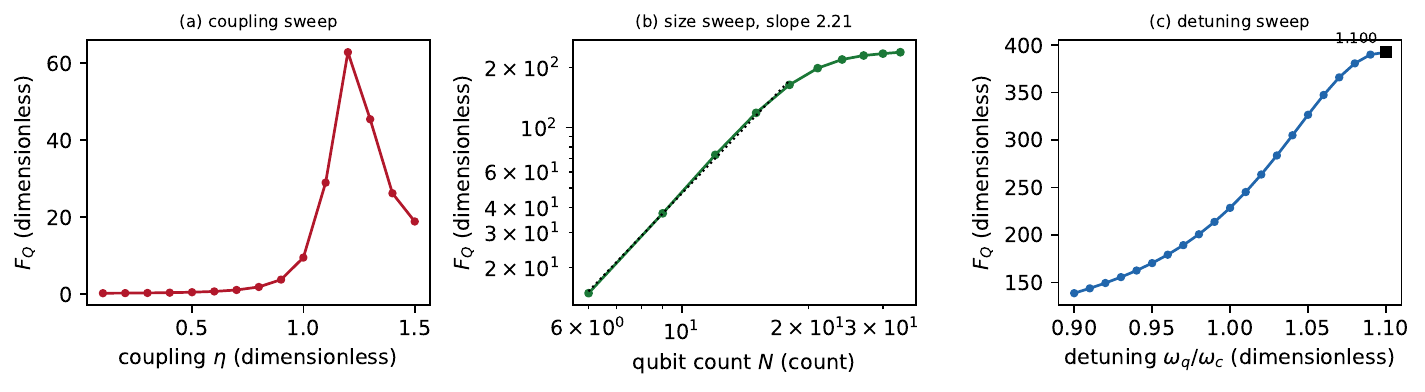}
\caption{Parameter sensitivity, all panels at $N=\num{32}$ and $n_{\max}=\num{35}$ except where stated. (a) QFI against coupling $\eta$ over $\eta\in\numrange{0.1}{1.5}$: the finite-size superradiant regime has a sharp onset and then flattens, with about \qty{7}{\percent} variation above $\eta=\num{1.1}$. (b) QFI against qubit count $N$ at $\eta=\num{1.15}$, $\omega_q/\omega_c=\num{1.005}$: a fit on $N\le\num{18}$ gives $F_Q\propto N^{\num{2.21}}$, which interpolates $N=\num{16}$ at about $\num{134}\times$. (c) QFI against detuning at the same size: maximum at $\omega_q/\omega_c=\num{1.10}$.}
\label{fig:si_sensitivity}
\end{figure}

\Cref{fig:si_sensitivity} resolves the parameter dependence behind those operating points: the superradiant onset in $\eta$, the collective exponent in $N$, and the detuning maximum at fixed size.

\section{Validity regimes and modeling scope}
\label{sec:si_scope_B}

All results in this Supplemental Material hold within the finite-dimensional Hilbert truncations, fixed-detuning Floquet drive conventions, two-level emitter spectra ($g_d/g_0\le\num{0.06}$ for multilevel leakage below \qty{1}{\percent} of the population), and Markovian Lindblad cavity-loss model analyzed here (with the \qty{4.8}{\percent}--\qty{24}{\percent} HEOM band bounding cavity-bath memory at physical linewidth, and incomplete Dicke+DCE basis convergence across $N_{\rm cav}=\numrange{6}{24}$). \Cref{tab:si_scope_matrix} summarizes each modeling assumption alongside its quantitative operational boundary.

\begin{table}[htbp]
\centering
\caption{Validity regimes, physical approximations, and quantitative operational boundaries constraining the ultrastrong-coupling metrology analysis.}
\label{tab:si_scope_matrix}
\footnotesize
\setlength{\tabcolsep}{4pt}
\begin{tabular}{l l l}
\toprule
    \parbox[t]{2.5cm}{\raggedright Physical dimension} & \parbox[t]{5.0cm}{\raggedright Modeled assumption} & \parbox[t]{6.6cm}{\raggedright Quantitative boundary / consequence} \\
\midrule
    \parbox[t]{2.5cm}{\raggedright Dicke coupling uniformity} & \parbox[t]{5.0cm}{\raggedright Identical $g_i=g$, no $\hat{\sigma}_+^{(i)}\hat{\sigma}_-^{(j)}$} & \parbox[t]{6.6cm}{\raggedright Inhomogeneous $g_i$ and dipolar crosstalk reduce collective gain} \\
    \parbox[t]{2.5cm}{\raggedright Floquet modulation} & \parbox[t]{5.0cm}{\raggedright Continuous-wave $g(t)$ without shaping} & \parbox[t]{6.6cm}{\raggedright Amplitude noise and gauge consistency~\cite{Stokes2021} require separate control} \\
    \parbox[t]{2.5cm}{\raggedright Multilevel emitter leakage} & \parbox[t]{5.0cm}{\raggedright Two-level truncation ($d_e=2$)} & \parbox[t]{6.6cm}{\raggedright Exceeds the \qty{1}{\percent} leak ceiling at $g_d/g_0=\num{0.11923}$; safe $g_d\le\num{0.06}$ gives \num{54}--\num{86}$\times$} \\
    \parbox[t]{2.5cm}{\raggedright Cavity decay channel} & \parbox[t]{5.0cm}{\raggedright Lindblad loss ($\sqrt{\kappa}\,\hat{a}$)} & \parbox[t]{6.6cm}{\raggedright Single-qubit retention $<\!\qty{1}{\percent}$ at $\kappa/2\pi=\qty{246}{\mega\hertz}$; Dicke+DCE $\sim\qty{3}{\percent}$--\qty{16}{\percent}} \\
    \parbox[t]{2.5cm}{\raggedright Qubit vs.\ cavity decoherence} & \parbox[t]{5.0cm}{\raggedright $T_1=\qty{25.8}{\micro\second}$, $T_\phi=\qty{111}{\micro\second}$ (computed for the circuit)} & \parbox[t]{6.6cm}{\raggedright Retains \qty{99.8}{\percent} alone vs.\ \qty{0.962}{\percent} with physical $\kappa$: cavity decay dominates} \\
    \parbox[t]{2.5cm}{\raggedright Cavity Fock truncation} & \parbox[t]{5.0cm}{\raggedright Finite cutoff $N_{\rm cav}=\numrange{6}{24}$} & \parbox[t]{6.6cm}{\raggedright Dicke+DCE retention drops (\qty{16}{\percent} $\to$ \qty{2.6}{\percent}); not cutoff-converged} \\
\bottomrule
\end{tabular}
\end{table}

\section{Empirical extraction fit and residual structure}
\label{sec:si_extraction}

The empirical extraction law is the quantitative summary of the retained-fraction collapse described in
\cref{MAIN-sec:extraction}. That it is fitted to the computed rows, rather than imposed as a transmissivity,
is what makes it an empirical statement about the collapse: the decay is not a single-scale process.

The empirical extraction law $R_Q=R_0[1+(\kappa\tau_{\rm int}/x_c)^{\beta}]^{-1}$ is fitted
to the \num{18} fitted rows of the single-qubit and $N=\num{32}$ Dicke
families, with $\tau_{\rm int}$ fixed row by row at the maximum of the retention
over the interrogation window reported in \cref{sec:si_interrogation}. The fit
gives $R_0=\num{0.9959}$, $x_c=\num{0.1271}$ with bootstrap \qty{95}{\percent}
interval $\numrange{0.107}{0.205}$, $\beta=\num{1.3816}$ with interval
$\numrange{1.019}{2.132}$, and $R^{2}=\num{0.9186}$. Both parameter intervals
are interior to the admissible box, so the exponent is a fitted quantity
here and not a bound of the search domain. Repeated with the saturation pinned
to the data-processing value $R_0=\num{1}$, the fit returns $x_c=\num{0.1260}$
with interval $\numrange{0.103}{0.190}$, $\beta=\num{1.3714}$ with interval
$\numrange{1.039}{1.931}$, and $R^{2}=\num{0.9186}$ against $\num{0.9186}$ for
the free fit. Two conclusions follow: the knee position and the exponent agree
across both fits to within their intervals, so neither answer depends on the
saturation being free. This stability reflects sampling the single-qubit crossover on both sides across $\kappa\tau_{\rm int}\in[\num{0.024},\num{0.18}]$, which pins $\beta$ to an interior minimum rather than the search boundary. The saturation $R_0=\num{0.9959}$ is unity to within the
reach of the fit, so the degradation of the extractable fraction is carried by
the knee position and the exponent rather than by a deficit at zero
accumulated loss.

The residual structure describes this finite set only. Across the eighteen fitted rows, retention remains below the bare transmissivity $\exp(-\kappa\tau_{\rm int})$; the smallest observed ratio is \num{0.0422}. In this model, the remaining difference is associated with qubit coherence and the shortened optimum interrogation window. The three rows with $\kappa\tau_{\rm int}<\num{0.01}$ have retentions \num{0.991}, \num{0.970} and \num{0.943}; the first sits within \qty{0.2}{\percent} of the floor set by the lossless row, so these residuals are dominated by estimator resolution rather than by a physical channel.

\subsection{Per-family knee and exponent}
\label{sec:si_extraction_families}

The pooled fit mixes two probe families, and the same two-parameter form fitted inside each family separates them (\cref{tab:si_extraction_families}). The single-qubit rows give $x_c=\num{0.109}$ with bootstrap \qty{95}{\percent} interval $\numrange{0.095}{0.117}$ and $\beta=\num{1.7336}$ with interval $\numrange{1.454}{2.035}$ ($R^{2}=\num{0.9928}$), and leave the saturation free at $R_0=\num{0.959}$ with $x_c=\num{0.117}$, $\beta=\num{1.902}$, $R^{2}=\num{0.9951}$. The four $N=\num{32}$ rows give $x_c=\num{0.371}$ with an interval $\numrange{0.174}{4.62}$ that runs past the sampled loss, and $\beta=\num{2.167}$ with interval $\numrange{0.606}{2.202}$ ($R^{2}=\num{0.9959}$); four rows cannot localize a knee, so only the pooled value should be read as a knee position. The two families therefore share the functional form but not a common scaling variable: pooling them costs $R^{2}=\num{0.919}$ against at least \num{0.99} inside each family, and the single-qubit knee lies outside the collective interval and conversely. What the collapse supports is that a single saturating curve brackets both families to within a factor \num{3} in $x_c$, not that one knee governs the two probe architectures.

\begin{table}[hbt!]
\centering
\caption{Empirical extraction fit $R_Q=R_0[1+(x/x_c)^{\beta}]^{-1}$ inside each probe family and on the pooled rows. Intervals are bootstrap \qty{95}{\percent} percentile intervals over \num{1000} resamples. The saturation is pinned to $R_0=\num{1}$ except in the last column of the single-qubit row, where it is free. $x_c$ and $\beta$ are dimensionless.}
\label{tab:si_extraction_families}
\begin{tabular}{l c c c c c c}
\toprule
Rows & $n$ & $x_c$ & \qty{95}{\percent} interval & $\beta$ & \qty{95}{\percent} interval & $R^{2}$ \\
\midrule
Single qubit & \num{14} & \num{0.1088} & \numrange{0.095}{0.117} & \num{1.7336} & \numrange{1.454}{2.035} & \num{0.9928} \\
$N=\num{32}$ Dicke & \num{4} & \num{0.3715} & \numrange{0.174}{4.62} & \num{2.1671} & \numrange{0.606}{2.202} & \num{0.9959} \\
Pooled & \num{18} & \num{0.1260} & \numrange{0.103}{0.190} & \num{1.3714} & \numrange{1.039}{1.931} & \num{0.9186} \\
\bottomrule
\end{tabular}
\end{table}

\subsection{Model selection and held-out error}
\label{sec:si_extraction_selection}

The word ``law'' is earned by ranking the saturating form against three alternatives on the same eighteen rows (\cref{tab:si_extraction_models}): the bare transmissivity $\exp(-x)$, a single exponential $A\exp(-x/x_0)$, and the three-parameter variant with a free saturation. The Gaussian AIC and BIC use the maximum-likelihood residual variance, and leave-one-out error is obtained by refitting on seventeen rows and predicting the held-out row, not by a leverage formula, so the nonlinearity of the knee is respected. The saturating form with pinned saturation has the lowest AIC; a single exponential trails it by $\Delta\mathrm{AIC}=\num{1.22}$ and the three-parameter variant by $\Delta\mathrm{AIC}=\num{1.99}$, while the bare transmissivity trails by $\Delta\mathrm{AIC}=\num{36.5}$. Leave-one-out RMSE orders the four forms identically, $\num{0.129}$ against $\num{0.131}$, $\num{0.131}$ and $\num{0.333}$. Two limits on this ranking should be stated. The margin over a single exponential is below the conventional $\Delta\mathrm{AIC}=\num{2}$ threshold, so the data prefer a saturating description over a bare exponential without excluding the exponential. And the held-out error is dominated by the two rows at the largest accumulated loss, where retention is near \qty{1}{\percent}; absolute residuals there reach \num{0.13}, so the criteria rank the forms on how they extrapolate into the collapsed regime rather than on their accuracy near the knee.

\begin{table}[hbt!]
\centering
\caption{Candidate forms for the retention collapse, fitted to the eighteen positive-loss rows. $k$ counts free parameters (dimensionless); $\mathrm{RSS}$ is the residual sum of squares (dimensionless); $\Delta\mathrm{AIC}$ is measured from the best form; leave-one-out $\mathrm{RMSE}$ is the held-out root-mean-square residual (dimensionless), equal to the in-sample value for the parameter-free transmissivity.}
\label{tab:si_extraction_models}
\begin{tabular}{l c c c c c}
\toprule
Form & $k$ & $\mathrm{RSS}$ & $\mathrm{AIC}$ & $\Delta\mathrm{AIC}$ & LOO $\mathrm{RMSE}$ \\
\midrule
Saturating, $R_0=1$ pinned & \num{2} & \num{0.2102} & \num{-76.10} & \num{0.00} & \num{0.1294} \\
Saturating, $R_0$ free & \num{3} & \num{0.2101} & \num{-74.11} & \num{1.99} & \num{0.1308} \\
Single exponential & \num{2} & \num{0.2249} & \num{-74.89} & \num{1.22} & \num{0.1312} \\
Bare transmissivity $e^{-x}$ & \num{0} & \num{1.9933} & \num{-39.61} & \num{36.49} & \num{0.3328} \\
\bottomrule
\end{tabular}
\end{table}

The largest computed cavity bases give physical-linewidth retentions of \qty{4.44}{\percent}, \qty{2.69}{\percent}, and \qty{2.58}{\percent} for $N_{\rm cav}=\num{16},\num{20},\num{24}$; this monotonic drop leaves complete-basis convergence beyond $N_{\rm cav}=\num{24}$ open for larger-basis studies.

\section*{Data availability}

The numerical data underlying every reported value, the simulation scripts and the environment specifications are available from the corresponding author on reasonable request.

\bibliographystyle{apsrev4-2}
\bibliography{04_track_B_metrology_references}